\documentclass[pra,aps,twocolumn]{revtex4}

\usepackage{amsmath}
\usepackage{epsfig}

\begin{document}
\title{Classification of multipartite entanglement containing
infinitely many kinds of states}
\author{Lin Chen}\email{deteriorate@zju.edu.cn}
\author{Yi-Xin Chen}\email{yxchen@zimp.edu.cn}
\author{Yu-Xue Mei}\email{deteriorate@zju.edu.cn}
\affiliation{Zhejiang Insitute of Modern Physics, Zhejiang
University, Hangzhou 310027, China}

\begin{abstract}
We give a further investigation of the range criterion and
Low-to-High Rank Generating Mode (LHRGM) introduced in
\cite{Chen}, which can be used for the classification of
$2\times{M}\times{N}$ states under reversible local filtering
operations. By using of these techniques, we entirely classify the
family of $2\times4\times4$ states, which actually contains
infinitely many kinds of states. The classifications of true
entanglement of $2\times(M+3)\times(2M+3)$ and
$2\times(M+4)\times(2M+4)$ systems are briefly listed
respectively.

\end{abstract}
\maketitle

\section{Introduction}
The rapid development of quantum information theory (QIT) requires
a further understanding of the properties of entanglement
\cite{Einstein}. Among many fundamental questions in QIT, it is
essential to find out how many different ways there exist, in
which several spatially distributed objects could be entangled
under certain prior constraint of physical resource, e.g., local
operations and classical communications (LOCC). This issue was
first addressed by Bennett \textit{et al.} \cite{Bennett1}. They
proved that the Bell pair
$(\left|00\right\rangle+\left|11\right\rangle)/\sqrt2$ is unique
in the pure settings when infinitely many copies of states are
available since the equivalent states can be used for the same
tasks in QIT.

The situation becomes complicated when only a single copy of state
is given. Due to the celebrated result by \cite{Lo}, the bipartite
pure states have been endowed with a nice classification under
LOCC. Unfortunately, there does not exist a similar theory in the
multipartite setting because the Schmidt polar form
\cite{Schmidt,Thapliyal,Carteret1} no longer acts here. In
addition, it turned out that the LOCC classification is lowly
efficient for the multipartite entanglement \cite{Linden1,Acin1}.
For simplicity, D\"ur \textit{et al.} \cite{Dur,Bennett2} has
considered the LOCC classification just in a stochastic manner
(SLOCC for short \cite{Gisin,Verstraete2}, or local filtering
operations), under which two states are equivalent if and only if
(iff) they are interconvertible with a nonvanishing probability.
By this criterion, they have explicitly shown that there exist two
sorts of fully entangled states in the three-qubit space, the GHZ
state $(\left|000\right\rangle+\left|111\right\rangle)/\sqrt2$ and
the W state
$(\left|001\right\rangle+\left|010\right\rangle+\left|100\right\rangle)/\sqrt3$
\cite{Bouwmeester}, which implies that the GHZ state is not the
sole representative inferred as before. Notwithstanding, the
existing results \cite{Verstraete1,Miyake2,Miyake3} concentrating
on the system with low dimensions showed that the SLOCC
classification is not a universally valid method for the
multipartite entanglement. On the other hand, it is of importance
to describe the structure of multipartite entanglement with higher
dimensions. This issue has been addressed in our earlier work
\cite{Chen}, where we have introduced the $range$ $criterion$ to
judge whether two multiple entangled states are equivalent under
SLOCC. Based on this criterion, we proposed the so-called
Low-to-High Rank Generating Mode (LHRGM) for the classification of
$2\times{M}\times{N}$ states. Specifically, we described how to
write out the essential classes of true entanglement in the
$2\times3\times3$, $2\times(M+1)\times(2M+1)$ and
$2\times(M+2)\times(2M+2)$ spaces respectively.

The main intention of this paper is to classify the family of
$2\times4\times4$ states, which actually contains infinitely many
entangled classes under SLOCC (we shall use $\sim$ to denote the
equivalence under SLOCC in this paper and only concern the true
entangled states whose local ranks do not change under SLOCC
\cite{Dur}). This helps analyze the structure of generally
multipartite entanglement which usually contains parameters. In
order to do this, we first make a further study of the techniques
including the range criterion and LHRGM in section II. In
particular, we will clarify that the range criterion is an
effective method for distinguishing the multipartite entangled
states, especially for those owning product states in some ranges
of the reduced density operators of these states. We also
exemplify its use in the more general cases such as the
$3\times3\times3$ and 4-qubit systems. As for the technique of
LHRGM, we focus on the analysis of the family of
$\left|\Omega_1\right\rangle$ which is a special branch in this
method. The result shows that this branch concerns the
classification of $2\times M\times M$ states. In fact, we discuss
here the classification of true entanglement with infinitely many
kinds of states under SLOCC related with $2\times M\times M$
states. Subsequently, we completely classify the $2\times4\times4$
states by LHRGM in section III. In section IV, we summarize the
existing results and propose the so-called quasi-combinatorial
character in any sequence of true $2\times M\times N$ systems,
$N=1,2,...,2M,$ and briefly give the classifications of true
entanglement of $2\times(M+3)\times(2M+3)$ and
$2\times(M+4)\times(2M+4)$ systems respectively, whose detailed
proofs are omitted. The conclusions are proposed in section V.

\section{some results from the range criterion and LHRGM}

Let us firstly recall the range criterion and the method of LHRGM
for the $2\times{M}\times{N}$ states, and the proof of these
techniques can be found in \cite{Chen}. A general pure state can
be expressed as $\rho_{A_1 A_2 \cdots
A_N}=\left|\Psi\right\rangle_{{A_1}{A_2}\cdots
{A_N}}\left\langle\Psi\right|$, and the reduced density operator
$\rho^{A_{ik+1},A_{ik+2},\cdots,A_{iN}}_\Psi\equiv$
tr$_{A_{i1},A_{i2},\cdots,A_{ik}}(\rho_{_{\scriptstyle \Psi}}),
i_1,i_2,\cdots,i_k\in\{1,2,\cdots,N\}, k\leq{N-1}$. Any state
$\left|\Psi\right\rangle_{{A_1}{A_2}\cdots {A_N}}$ in the
$D_1\times D_2\times\cdots\times D_N$ space ( sometimes also
written
$\left|\Psi\right\rangle_{{D_1}\times{D_2}\times\cdots\times{D_N}}$
) can always be transformed into the adjoint form,
\begin{equation}
\left|\Phi\right\rangle=\sum_{i=0}^{D_j-1}
\left|i\right\rangle_{A_j}\otimes\left|i\right\rangle_{{A_1}{A_2}\cdots{A_{j-1}}{A_{j+1}}\cdots{A_N}},
\end{equation}
where the computational basis
$\langle{i}|{k}\rangle_{A_j}={\delta}_{ik}$ and
$\{\left|i\right\rangle_{{A_1}{A_2}\cdots{A_{j-1}}{A_{j+1}}\cdots{A_N}},i=0,1,\cdots,D_j-1\}$
are a set of linearly independent vectors, each
$\left|i\right\rangle_{{A_1}{A_2}\cdots{A_{j-1}}{A_{j+1}}\cdots{A_N}}$
is the adjoint state of $\left|i\right\rangle_{A_j}$. If $\rho$
acts on the Hilbert space $\cal H$, then the range of $\rho$ is
$R(\rho)=\rho\left|\Phi\right\rangle,\left|\Phi\right\rangle\in\cal
H$ \cite{Horn}. Then we can write out the range criterion as
follows.

\textit{Range Criterion}. For two multiple states
$\left|\Psi\right\rangle_{{A_1}{A_2}\cdots{A_N}}$ and
$\left|\Phi\right\rangle_{{A_1}{A_2}\cdots{A_N}}$, there exist
certain ILO's $V_i, i=1,...,N$ making
$\left|\Psi\right\rangle_{{A_1}{A_2}\cdots{A_N}}=
{V_1}\otimes{V_2}\otimes\cdots\otimes{V_N}\left|\Phi\right\rangle_{{A_1}{A_2}\cdots{A_N}}$
iff there exist a series of numbers
$i_1,i_2,\cdots,i_{N-1}\in\{1,2,\cdots,N\},$ such that $
R(\rho^{A_{i1},A_{i2},\cdots,A_{i_{N-1}}}_\Psi)={V_{i1}}\otimes{V_{i2}}\otimes\cdots\otimes{V_{i_{N-1}}}
R(\rho^{A_{i1},A_{i2},\cdots,A_{i_{N-1}}}_\Phi)$. Let
$[a_1,a_2,a_3,...,a_N]$ represent a set of states, and a state
$\left|\Phi\right\rangle_{{A_1}{A_2}\cdots{A_N}}\in[a_1,a_2,a_3,...,a_N]$
iff the number of product states in
$R(\rho^{{A_1}\cdots{A_{i-1}}{A_{i+1}}\cdots{A_N}}_{\Phi_{{A_1}{A_2}\cdots{A_N}}})$
is $a_i$, $i=2,...,N-1$.

The range criterion has been used for the classification of
several sorts of true entanglement in \cite{Chen}, and here we
would like to further discuss how to generically distinguish the
triple entangled states in $2\times M\times N$ space. For triple
qubit states, a method of practical identification has been given
in \cite{Dur}, where the 3-tangle \cite{Coffman} (decided by
concurrence \cite{Wootters}) is employed as a criterion. Recently,
resultful progress in calculation of the concurrence of arbitrary
bipartite states has been achieved by \cite{Kai}, but it is
unclear that whether there exist a generally certain criterion
determining the relation between a multiple entangled state and
the concurrences of its reduced density operators in all bipartite
subspaces, simply similar to that in \cite{Dur}. In fact, it
remains a formidable challenge to QIT \cite{Mintert}. While in
\cite{Chen}, the range criterion provides a universal and
effective method to distinguish the multipartite entanglement in
$2\times M\times N$ space. We summarize this procedure as follows.
Given a set of states $\left|\psi_i\right\rangle,i=1,2,...,$ in
the $2\times M\times N$ space. One first finds out their adjoint
forms respectively. Then the set can be split into several subsets
of entangled states with different local ranks. Clearly, any two
states from different subsets are inequivalent under SLOCC
\cite{Dur}. For each subset, one applies the range criterion to
judge whether the states are equivalent. The concrete example has
been given in \cite{Chen}, where the product states in the ranges
greatly help the judgement. Moreover, one can find out the ILO's
between two equivalent states through a procedure similar to the
arguments in \cite{Chen}, while the calculation therein is
relatively succinct.

It should be noted that there exist a special kind of state, i.e.,
$\left|\Psi\right\rangle_{2\times M\times M}$. For this case, one
has to compare whether
$\left|\Psi\right\rangle_{ABC}\sim\left|\Psi\right\rangle_{ACB}$.
One may guess that the above relation always holds, for the
existing results for $2\times2\times2$ and $2\times3\times3$
systems support it. However as shown in the next section, it is
not true in the case of $2\times4\times4$ system \cite{notation1}.
In general, a difficult problem emerges when trying to classifying
the multiple entangled states, i.e., the permutation of the
parties makes the situation much more sophisticated, when (part
of) the local ranks of the state are identical. For example, the
multiqubit system can be in distinct state by exchanging the
parties therein, which may be the most trouble finding out the
essential classes of multiqubit states. In this case, the range
criterion can effectively help analyze the structure of
multipartite entanglement.

For instance, consider a family of 4-qubit state
\begin{equation}
\left|\Phi_x\right\rangle_{ABCD}=\left|00\right\rangle(\left|00\right\rangle+\left|11\right\rangle)
+\left|11\right\rangle(\left|00\right\rangle+x\left|11\right\rangle),x\neq1.
\end{equation}
For simplicity we can generalize the definition of
$[a_1,a_2,a_3,...,a_N]$ in the range criterion, e.g., a state
$\left|\Phi\right\rangle_{ABCD}\in[a_1,a_2,a_3,a_4,a_5,a_6]$ iff
the number of product states in $R(\rho^{AB}_{\Phi_x})$ is $a_1$,
in $R(\rho^{AC}_{\Phi_x})$ is $a_2$, in $R(\rho^{AD}_{\Phi_x})$ is
$a_3$, in $R(\rho^{BC}_{\Phi_x})$ is $a_4$, in
$R(\rho^{BD}_{\Phi_x})$ is $a_5$ and in $R(\rho^{CD}_{\Phi_x})$ is
$a_6$. Thus we denote that
$\left|\Phi_x\right\rangle_{ABCD}\in[2,\infty,\infty,\infty,\infty,2]$.
Due to the range criterion, the possible equivalence by
permutation is nothing but
$\left|\Phi_x\right\rangle_{ABCD}\sim\left|\Phi_x\right\rangle_{CDAB}$
(the two subsystems AB, and CD are symmetric respectively). So we
can easily dispose of this 4-qubit family in terms of the range
criterion, and the result is
$\left|\Phi_x\right\rangle_{ABCD}\sim\left|\Phi_{1/x}\right\rangle_{ABCD}
\sim\left|\Phi_x\right\rangle_{CDAB}\sim\left|\Phi_{1/x}\right\rangle_{CDAB}$.
A more interesting result is about the 3-qutrit state
\begin{equation}
\left|\Psi\right\rangle_{ABC}=\left|001\right\rangle+\left|010\right\rangle+\left|100\right\rangle+
\left|220\right\rangle+\left|202\right\rangle\in[1,\infty,\infty].
\end{equation}
Now the exchange of the parties indeed leads to inequivalent
classes of entanglement,
\begin{equation}
\left|\Psi\right\rangle_{BAC}=\left|001\right\rangle+\left|010\right\rangle+\left|100\right\rangle+
\left|220\right\rangle+\left|022\right\rangle\in[\infty,1,\infty],
\end{equation}
and
\begin{equation}
\left|\Psi\right\rangle_{CBA}=\left|001\right\rangle+\left|010\right\rangle+\left|100\right\rangle+
\left|022\right\rangle+\left|202\right\rangle\in[\infty,\infty,1].
\end{equation}
Notice there is no other new classes derived from the permutation
of parties by symmetry. So the range criterion is often practical
to judge whether a state is $symmetric$ or the exchange of the
parties cause new classes of states whose local ranks are (partly)
identical.

Next, we explore the possibility applying the range criterion to
the classification of more universal entangled states. For the
systems with higher dimensions, one may still employ the range
criterion to distinguish the given states. However, the situations
therein are more complicated for there are not always product
states in the ranges of reduced density operators in bipartite
subspaces. For example, consider such a family of entangled states
in the $3\times3\times3$ space,
\begin{equation}
\left|\Psi_x\right\rangle_{ABC}=\left|012\right\rangle+\left|120\right\rangle
+\left|201\right\rangle+\left|000\right\rangle+\left|111\right\rangle+x\left|222\right\rangle,\\
\end{equation}
where the parameter $x\neq0$ cannot be removed by ILO's. One can
readily check that there is no product state in
$R(\rho^{AB}_{\Psi_x})$, $R(\rho^{AC}_{\Psi_x})$, and
$R(\rho^{BC}_{\Psi_x})$, for all adjoint states are of entangled
forms. It is then not easy to classify this family of states, we
briefly make out the clue to this case. Clearly,
$\left|\Psi_x\right\rangle_{ABC}$ is symmetric by the operations
$\left|0\right\rangle\leftrightarrow\left|1\right\rangle$ on each
party. A primary observation tells that
$\left|\Psi_x\right\rangle\sim\left|\Psi_{x^{-1}}\right\rangle$,
which can be realized by the ILO's
$V^{3\times3}_A=$diag$(1,1,x^{-1})$ and two permutation
transformations\\
$V_B^{3\times3}=\left(\begin{array}{cccc}
0 & 1 & 0 \\
0 & 0 & 1 \\
1 & 0 & 0
\end{array}\right),
V_C^{3\times3}=\left(\begin{array}{cccc}
0 & 0 & 1 \\
1 & 0 & 0 \\
0 & 1 & 0
\end{array}\right)$,\\
up to some permutation of the parties. In fact, there does not
exist other equivalence relations in this 3-qutrit family. To
simplify the proof, we adopt the concept of normal form of a pure
multipartite state \cite{Verstraete}. That is, one can transform
any $\left|\Phi\right\rangle_{{A_1}{A_2}\cdots{A_N}}$ into its
normal form by the ILO's, whose local density operators are all
proportional to the identity and the normal form is unique up to
local unitary transformations. In the light of this fact, by the
ILO's
$V^{3\times3}_A=V^{3\times3}_B=V^{3\times3}_C=$diag$(1,1,x^{-1/3})$,
we obtain the normal form of $\left|\Psi_x\right\rangle_{ABC}$,
\begin{eqnarray}
\left|\Psi^{norm}_x\right\rangle_{ABC}&=&x^{-1/3}\left|012\right\rangle+x^{-1/3}\left|120\right\rangle
+x^{-1/3}\left|201\right\rangle\nonumber\\
&+&\left|000\right\rangle+\left|111\right\rangle+\left|222\right\rangle,
\end{eqnarray}
whose local density operators are indeed proportional to the
identity, so the possible ILO's making
$\left|\Psi^{norm}_x\right\rangle_{ABC}$ into
$\left|\Psi^{norm}_y\right\rangle_{ABC}$ must be unitary.
Moreover, one can readily check that the local rank of the states
in $R(\rho^{AB}_{\Psi^{norm}_x})$ ( also
$R(\rho^{AC}_{\Psi^{norm}_x})$, and $R(\rho^{BC}_{\Psi^{norm}_x})$
) can be no more than two iff $x=-1$, while in any other case the
local rank of the combination of three adjoint states is always
three, so there exist at least one zero in every column of the
unitary transformations. Recall that a unitary matrix acting on a
$d-$dimensional space can always be written as a product of
several two-level unitary matrices \cite{Chuang}. With these
techniques one can find out the form of the unitary
transformations and get the above assertion. A striking feature of
the state $\left|\Psi_x\right\rangle_{ABC}$ is that it actually
contains $infinitely$ many classes of entanglement, which is the
main topic in this paper. Besides this example, we will
demonstrate the concrete techniques for such generic system by the
classification of $2\times4\times4$ states in next section. On the
other hand, intuitively the cases with no product states in the
ranges are more involved than those owning at least one product
state. Unfortunately, finding out the normal form of a state is
often difficult, or sometimes meaningless if it is identical to
zero \cite{Verstraete}. Although one can follow the formal
procedure similar to that in the appendix in \cite{Chen}, it is
now difficult to find a solution, or to prove there is no solution
for the equations set. So the range criterion works more
effectively when there is at least one product state in some range
of the reduced density operator in bipartite subspace.

Let us move to the method of Low-to-High Rank Generating Mode
(LHRGM), which can be expressed as follows.

\textit{LHRGM.} For the classification of true tripartite
entangled states under SLOCC, the following equivalence relation
is
true, \\
$ \left|\Psi\right\rangle_{2\times M \times N}\sim
\left\{\begin{array}{l}
\left|\Omega_0\right\rangle\equiv(a\left|0\right\rangle
+b\left|1\right\rangle)\left|M-1,N-1\right\rangle
\\+\left|\Psi\right\rangle_{2\times(M-1) \times(N-1)},\\
\left|\Omega_1\right\rangle\equiv\left|0,M-1,N-1\right\rangle
\\+\left|1,M-1,N-2\right\rangle+\left|\Psi\right\rangle_{2\times(M-1)\times(N-2)},\\
\left|\Omega_2\right\rangle\equiv\left|\Omega_0\right\rangle
+\left|0,M-1\right\rangle\left|\chi\right\rangle,b\neq0,\\
\left|\Omega_3\right\rangle\equiv\left|\Omega_0\right\rangle
+\left|1,M-1\right\rangle\left|\chi\right\rangle,a\neq0.
\end{array}\right.$\\
The condition $a\neq0$ or $b\neq0$ keeps
$\left|\Omega_2\right\rangle$ and $\left|\Omega_3\right\rangle$
not becoming $\left|\Omega_0\right\rangle$.
$\left|\chi\right\rangle\equiv\sum_{i=0}^{N-2}
a_i\left|i\right\rangle$ and the arbitrary constants $a_i$'s do
not equal zero simultaneously.

Notice that all related states $\left|\Psi\right\rangle_{2\times M
\times N}$ in the \textit{LHRGM} are truly entangled. The LHRGM is
indeed an iterated method, in the sense that the calculation of
the essential classes of highly dimensional states requires that
of low-level families. As shown in \cite{Chen}, one first
calculates $\left|\Omega_0\right\rangle$, which requires a
relatively low amount of calculations. Based on it, the
$\left|\Omega_2\right\rangle$ and $\left|\Omega_3\right\rangle$'s
family can be calculated economically, where several important
invariance of ILO's have been employed so that the calculation can
be enormously simplified. The calculation of above three families
are necessarily required for any classification of
$\left|\Psi\right\rangle_{2\times M \times N}$. Moreover, in the
cases of $2\times(M+1)\times(2M+1)$ and $2\times(M+2)\times(2M+2)$
system in \cite{Chen}, the $\left|\Omega_1\right\rangle$'s family
proved to be a subset of the above three families by induction.

We advance another problem on this special family. That is, does
any $\left|\Omega_1\right\rangle$'s family always belong to
corresponding $\left|\Omega_0\right\rangle$,
$\left|\Omega_2\right\rangle$ and $\left|\Omega_3\right\rangle$'s
family? The answer is negative, since we have found an exception
in the $2\times3\times3$ system, i.e., the state
$\left|001\right\rangle+\left|010\right\rangle+\left|112\right\rangle+\left|112\right\rangle$.
By theorem 2 in \cite{Chen}, it does not belong to anyone of other
three families, which indicates that the induction will not be
available for every classification of entanglement in $2\times
M\times N$ space. Apply the LHRGM technique to the term
$\left|\Psi\right\rangle_{2\times(M-1)\times(N-2)}$ in
$\left|\Omega_1\right\rangle$,\\
$\left|\Omega_1\right\rangle_{2\times M \times
N}\sim\left|0,M-1,N-1\right\rangle +\left|1,M-1,N-2\right\rangle+
\left\{\begin{array}{l}
(a\left|0\right\rangle+b\left|1\right\rangle)\left|M-2,N-3\right\rangle\\
+\left|\Psi\right\rangle_{2\times(M-2)\times(N-3)},\hspace*{\fill}(I)\\
\left|0,M-2,N-3\right\rangle\\
+\left|1,M-2,N-4\right\rangle+\left|\Psi\right\rangle_{2\times(M-2)\times(N-4)},\hspace*{\fill}(II)\\
(a\left|0\right\rangle+b\left|1\right\rangle)\left|M-2,N-3\right\rangle\\
+\left|0,M-2\right\rangle\left|\chi^{\prime}\right\rangle
+\left|\Psi\right\rangle_{2\times(M-2)\times(N-3)},b\neq0,\hspace*{\fill}(III)\\
(a\left|0\right\rangle+b\left|1\right\rangle)\left|M-2,N-3\right\rangle\\
+\left|1,M-2\right\rangle\left|\chi^{\prime}\right\rangle
+\left|\Psi\right\rangle_{2\times(M-2)\times(N-3)},a\neq0,\hspace*{\fill}(IV)
\end{array}\right.$\\
where
$\left|\chi^{\prime}\right\rangle=\sum_{i=0}^{N-4}a_i\left|i\right\rangle$.
As for expression (I), one performs
$\left|M-1\right\rangle_B\leftrightarrow\left|M-2\right\rangle_B$
and
$\left|N-1\right\rangle_C\leftrightarrow\left|N-3\right\rangle_C$,
so that $(I)\sim\left|\Omega_0\right\rangle_{2\times M \times N}$,
which becomes one of other three families. The same ILO's make
$(III)$ and $(IV)$ identical to
$\left|\Omega_2\right\rangle_{2\times M \times N}$ or
$\left|\Omega_3\right\rangle_{2\times M \times N}$. Thus, the only
exception is $(II)\sim\left|0,M-1,N-1\right\rangle
+\left|1,M-1,N-2\right\rangle+\left|0,M-2,N-3\right\rangle
+\left|1,M-2,N-4\right\rangle+\left|\Psi\right\rangle_{2\times(M-2)\times(N-4)}$.
Continue this procedure, which will cease when $M-1-k=N-2-2k$, or
$k=N-M-1$. Thus the unique exception is
\begin{eqnarray}
\left|\Psi\right\rangle_{2\times M\times
N}^{exc}\equiv\sum_{i=0}^{N-M-1}(\left|0,M-1-i,N-1-2i\right\rangle+ \nonumber\\
\left|1,M-1-i,N-2-2i\right\rangle)+\left|\Psi\right\rangle_{2\times(2M-N)\times(2M-N)}^{\prime}.
\end{eqnarray}
Notice there are no two identical terms in
$\{\left|N-1-2i\right\rangle,\left|N-2-2i\right\rangle\},i=0,1,...,N-M-1$.
We investigate the above expression in several cases. ( the states
$\left|\Upsilon_i\right\rangle,i=0,1,2$ are defined in
\cite{Chen}, their explicit forms are given in the section IV )
\\(i) $N=2M.$ The state $\left|\Psi\right\rangle_{2\times M\times
2M}^{exc}$ is just the unique class
$\left|\Upsilon_0\right\rangle$ in $2\times M\times 2M$ space. \\
(ii) $N=2M-1.$ It is the case of theorem 3 in \cite{Chen}.
Evidently, we have $\left|\Psi\right\rangle_{2\times M\times
(2M-1)}^{exc}\sim\left|\Upsilon_1\right\rangle$, which belongs to
the $\left|\Omega_0\right\rangle$'s family.\\
(iii) $N=2M-2.$ It is the case of theorem 4 in \cite{Chen}. There
are two situations here corresponding to $\left|GHZ\right\rangle$
and $\left|W\right\rangle$ state respectively. If
$\left|\Psi\right\rangle_{2\times2\times2}^{\prime}\sim\left|GHZ\right\rangle$,
by operations
$\left|N-1\right\rangle_C\leftrightarrow\left|0\right\rangle_C$
and
$\left|M-1\right\rangle_B\leftrightarrow\left|0\right\rangle_B$,
we then transform $\left|\Psi\right\rangle_{2\times
M\times(2M-2)}^{exc}$ into $\left|\Omega_0\right\rangle$'s family.
On the other hand if
$\left|\Psi\right\rangle_{2\times2\times2}^{\prime}\sim\left|W\right\rangle$,
by the same ILO's $\left|\Psi\right\rangle_{2\times
M\times(2M-2)}^{exc}$ always becomes either
$\left|\Omega_2\right\rangle$ or $\left|\Omega_3\right\rangle$.
Therefore, (ii) and (iii) confirm the fact that
$\left|\Omega_1\right\rangle_{2\times(M+1)\times(2M+1)}$ and
$\left|\Omega_1\right\rangle_{2\times(M+2)\times(2M+2)}$ belong to
other three relevant families respectively. \\
(iv) $N=2M-3$, and we get
$\left|\Psi\right\rangle_{2\times3\times3}^{\prime}$. By the
result of $2\times3\times3$ states in \cite{Chen} and skills
similar to that in (iii), we find the unique exception
\begin{eqnarray*}
\left|\Psi\right\rangle_{2\times M\times
(2M-3)}^{exc}=\sum_{i=0}^{M-4}(\left|0,M-1-i,2M-4-2i\right\rangle+\nonumber\\
\left|1,M-1-i,2M-5-2i\right\rangle)
+\left|010\right\rangle+\left|001\right\rangle+
\left|112\right\rangle+\left|121\right\rangle.\nonumber\\
\end{eqnarray*}
Thus we only need to calculate other three families when we
classify the entanglement in $2\times M\times(2M-3)$ space. In
general, it is necessary to analyze the expression of
$\left|\Psi\right\rangle_{2\times M\times N}^{exc}$ since the
induction operates merely in the cases of (ii) and (iii).
Conclusively, the LHRGM technique equals the calculation of
$\left|\Omega_0\right\rangle$, $\left|\Omega_2\right\rangle$,
$\left|\Omega_3\right\rangle$ and
$\left|\Psi\right\rangle_{2\times M\times N}^{exc}$.

So far we have managed to check several special cases of
$\left|\Psi\right\rangle_{2\times M\times N}^{exc}$. Evidently, it
is determined by
$\left|\Psi\right\rangle_{2\times(2M-N)\times(2M-N)}^{\prime}$
that whether $\left|\Omega_1\right\rangle_{2\times M\times N}$
will produce exceptional states not belonging to the other
families. Let $Q=2M-N$. As for the case of $Q=3$, we have found
the essential classes of
$\left|\Psi\right\rangle_{2\times3\times3}$ in \cite{Chen}. With
the help of existing results and techniques, one can step-by-step
calculate each $\left|\Psi\right\rangle_{2\times Q\times
Q}^{\prime}$'s family by LHRGM, $Q=4,5,...$. It is difficult to
provide a restrict criterion determining whether the exceptional
family are equivalent to those derived from
$\left|\Omega_0\right\rangle$, $\left|\Omega_2\right\rangle$ and
$\left|\Omega_3\right\rangle$, for one has to use the LHRGM
technique in the $ABC$ and $ACB$ system in turn, since there will
exist the cases of $M>N$. In addition, there will more frequently
be of infinitely many classes of entanglement in this family (see
next section), which also enhances its complexity. On all
accounts, the calculation of essential classes of
$\left|\Psi\right\rangle_{2\times M\times(2M-Q)}$ by LHRGM
requires that one knows the classes of true entangled states of
$\left|\Psi\right\rangle_{2\times (M-1)\times(2M-Q-1)}$ and
$\left|\Psi\right\rangle_{2\times Q\times Q}$.

To summary, we have reviewed the range criterion and applied it to
the more general cases such as the 4-qubit and 3-qutrit systems,
which can be entangled in infinitely many ways. For the technique
of LHRGM, we have analyzed the possible structure of
$\left|\Omega_1\right\rangle$ and the method of its simplified
calculation.

\section{classification of $2\times4\times4$ states}

As mentioned in the preceding section, a distributed system can be
entangled in infinitely many ways with increasing of dimensions of
Hilbert space. The simplest case is the family of
$2\times4\times4$ states, so it worth a further investigation. An
early result by W. D\"ur \textit{et al.} \cite{Dur}, has pointed
out that the entangled systems with a finite number of entangled
classes only potentially exist in the $2\times M\times N$ space.
This assertion has been confirmed in the four-qubit system
\cite{Verstraete1}. Here, we will prove that the family of
$2\times4\times4$ states also contains infinitely many essential
classes. In addition, the $2\times4\times4$ system can be regarded
as a special case of the five-qubit system. Hence, the
classification of $2\times4\times4$ states helps get insight into
the structure of the five-qubit states, which remains a
sophisticated problem in QIT.

Before we go to the classification of $2\times4\times4$ states, we
shall make some useful preparations , mainly quoted from
\cite{Chen}. Define two ILO's
$O^A_1(\left|\phi\right\rangle,\alpha):
\left|\phi\right\rangle_A\rightarrow\alpha\left|\phi\right\rangle_A$
and $O^A_2(\left|\phi\right\rangle,\left|\psi\right\rangle):
\left|\phi\right\rangle_A\rightarrow\left|\phi\right\rangle_A+\left|\psi\right\rangle_A$,
respectively. Some existing results are also available. They are
the $2\times3\times2$ classes of entanglement,
\begin{eqnarray}
\left|\phi_0\right\rangle&\equiv&\left|000\right\rangle+\left|011\right\rangle
+\left|121\right\rangle\in[1,\infty,1],\nonumber\\
\left|\phi_1\right\rangle&\equiv&\left|000\right\rangle+\left|011\right\rangle
+\left|110\right\rangle+\left|121\right\rangle\in[0,\infty,0],
\end{eqnarray}
and the $2\times3\times3$ classes,
\begin{eqnarray}
\left|\varphi_0\right\rangle&\equiv&\left|000\right\rangle+\left|111\right\rangle
+\left|022\right\rangle\in[1,\infty,\infty],\nonumber\\
\left|\varphi_1\right\rangle&\equiv&\left|000\right\rangle+\left|111\right\rangle
+(\left|0\right\rangle+\left|1\right\rangle)\left|22\right\rangle\in[0,3,3],\nonumber\\
\left|\varphi_2\right\rangle&\equiv&\left|010\right\rangle+\left|001\right\rangle
+\left|112\right\rangle+\left|121\right\rangle\in[0,\infty,\infty],\nonumber\\
\left|\varphi_3\right\rangle&\equiv&\left|100\right\rangle+\left|010\right\rangle
+\left|001\right\rangle+\left|112\right\rangle+\left|121\right\rangle\in[0,1,1],\nonumber\\
\left|\varphi_4\right\rangle&\equiv&\left|100\right\rangle+\left|010\right\rangle
+\left|001\right\rangle+\left|022\right\rangle\in[1,\infty,\infty],\nonumber\\
\left|\varphi_5\right\rangle&\equiv&\left|100\right\rangle+\left|010\right\rangle
+\left|001\right\rangle+\left|122\right\rangle\in[0,2,2].
\end{eqnarray}
All these states are inequivalent under SLOCC in terms of the
range criterion. Notice there exist the so-called invariance of
$\left|\varphi_1\right\rangle$ under the ILO's
$\left|0\right\rangle\leftrightarrow\left|1\right\rangle$ on all
parties, and that of $\left|\varphi_2\right\rangle$ under the
ILO's
$\left|0\right\rangle_A\leftrightarrow\left|1\right\rangle_A$,
$\left|0\right\rangle_B\leftrightarrow\left|2\right\rangle_B$ and
$\left|0\right\rangle_C\leftrightarrow\left|2\right\rangle_C$.
This feature will help simplify the calculation. With these
preconditions, we can describe the classification of
$2\times4\times4$ states.

\textit{Theorem 1.} There are 16 essentially entangled classes in
$2\times4\times4$ space,
\begin{eqnarray}
\left|\Phi_0\right\rangle&\equiv&\left|133\right\rangle+
\left|\varphi_0\right\rangle\in[0,\infty,\infty];\nonumber\\
\left|\Phi_1\right\rangle&\equiv&\left|033\right\rangle+
\left|\varphi_0\right\rangle\in[1,\infty,\infty];\nonumber\\
\left|\Phi_2\right\rangle&\equiv&\left|133\right\rangle+
\left|\varphi_1\right\rangle\in[0,\infty,\infty];\nonumber\\
\left|\Phi_3\right\rangle_x&\equiv&(\left|0\right\rangle+x\left|1\right\rangle)\left|33\right\rangle+
\left|\varphi_1\right\rangle\in[0,4,4],x\neq0,1;\nonumber\\
\left|\Phi_4\right\rangle&\equiv&\left|133\right\rangle+
\left|\varphi_2\right\rangle\in[0,\infty,\infty];\nonumber\\
\left|\Phi_5\right\rangle&\equiv&\left|133\right\rangle+
\left|\varphi_3\right\rangle\in[0,\infty,\infty];\nonumber\\
\left|\Phi_6\right\rangle&\equiv&\left|033\right\rangle+
\left|\varphi_3\right\rangle\in[0,2,2];\nonumber\\
\left|\Phi_7\right\rangle&\equiv&\left|133\right\rangle+
\left|\varphi_4\right\rangle\in[0,\infty,\infty];\nonumber\\
\left|\Phi_8\right\rangle&\equiv&\left|033\right\rangle+
\left|\varphi_4\right\rangle\in[1,\infty,\infty];\nonumber\\
\left|\Phi_9\right\rangle&\equiv&\left|133\right\rangle+
\left|\varphi_5\right\rangle\in[0,\infty,\infty];\nonumber\\
\left|\Phi_{10}\right\rangle&\equiv&(\left|0\right\rangle+\left|1\right\rangle)\left|33\right\rangle+
\left|\varphi_5\right\rangle\in[0,3,3];\nonumber\\
\left|\Phi_{11}\right\rangle&\equiv&\left|133\right\rangle+\left|032\right\rangle+
\left|\varphi_2\right\rangle\in[0,\infty,\infty];\nonumber\\
\left|\Phi_{12}\right\rangle&\equiv&\left|133\right\rangle+\left|032\right\rangle+
\left|\varphi_3\right\rangle\in[0,1,1];\nonumber\\
\left|\Phi_{13}\right\rangle&\equiv&\left|033\right\rangle+\left|132\right\rangle+
\left|\varphi_4\right\rangle\in[0,\infty,\infty];\nonumber\\
\left|\Phi_{14}\right\rangle&\equiv&\left|133\right\rangle+\left|032\right\rangle+
\left|\varphi_5\right\rangle\in[0,2,2];\nonumber\\
\left|\Phi_{15}\right\rangle&\equiv&\left|033\right\rangle+\left|132\right\rangle+
\left|\phi_1\right\rangle\in[0,\infty,\infty].
\end{eqnarray}
Notice the class $\left|\Phi_3\right\rangle$ contains a parameter
$x$, which cannot be removed. So the parameter $x$ characterizes
that the classes of true entangled states of $2\times4\times4$
system are infinite.

\textit{Proof.} Due to the LHRGM we write out \\
$\left|\Psi\right\rangle_{2\times4\times4}\sim
\left\{\begin{array}{l}
\left|\Omega_0\right\rangle=(a\left|0\right\rangle
+b\left|1\right\rangle)\left|33\right\rangle+\left|\Psi\right\rangle_{2\times3\times3},\\
\left|\Omega_1\right\rangle=\left|033\right\rangle
+\left|132\right\rangle+\left|\Psi\right\rangle_{2\times3\times2},\\
\left|\Omega_2\right\rangle=\left|\Omega_0\right\rangle
+\left|03\right\rangle\left|\chi\right\rangle,b\neq0,\\
\left|\Omega_3\right\rangle=\left|\Omega_0\right\rangle
+\left|13\right\rangle\left|\chi\right\rangle,a\neq0.
\end{array}\right.$\\
Here,
$\left|\chi\right\rangle=\sum^2_{i=0}a_i\left|i\right\rangle$
since $N=4$. Let us begin with the calculation of
$\left|\Omega_0\right\rangle$. The style is similar to \cite{Chen}.\\
(i) $\left|\Omega_0\right\rangle_0=(a\left|0\right\rangle
+b\left|1\right\rangle)\left|33\right\rangle+\left|\varphi_0\right\rangle$.
If $ab=0,$ then it leads to the entangled class
$\left|\Phi_0\right\rangle$ or $\left|\Phi_1\right\rangle$. For
the case of $ab\neq0,$ it holds that
$\left|\Omega_0\right\rangle_0\sim(\left|0\right\rangle
+\alpha\left|1\right\rangle)\left|33\right\rangle+\left|000\right\rangle+\left|111\right\rangle
+\left|022\right\rangle$. By operations
$O^A_1(\left|1\right\rangle,\alpha^{-1})\otimes
O^B_1(\left|1\right\rangle,\alpha)$,
$\left|2\right\rangle_B\leftrightarrow\left|3\right\rangle_B$,
$\left|2\right\rangle_C\leftrightarrow\left|3\right\rangle_C$ and
$\left|0\right\rangle\leftrightarrow\left|1\right\rangle$ in all
parties, we make $\left|\Omega_0\right\rangle_0$ go into the class $\left|\Phi_2\right\rangle$.\\
(ii) $\left|\Omega_0\right\rangle_1=(a\left|0\right\rangle
+b\left|1\right\rangle)\left|33\right\rangle+\left|\varphi_1\right\rangle$.
Due to the invariance of $\left|\varphi_1\right\rangle$, for the
case of $ab=0$ it is transformed into $\left|\Phi_2\right\rangle$.
If $ab\neq0,$ then
$\left|\Omega_0\right\rangle_1\sim(\left|0\right\rangle
+\alpha\left|1\right\rangle)\left|33\right\rangle+\left|000\right\rangle+\left|111\right\rangle
+(\left|0\right\rangle+\left|1\right\rangle)\left|22\right\rangle
\sim\left|\Phi_3\right\rangle_x,x\neq0,1$. If $x=1$
$\left|\Omega_0\right\rangle_1$ is taken into the class
$\left|\Phi_2\right\rangle$ by
$O^A_2(\left|1\right\rangle,-\left|0\right\rangle)\otimes
O^A_1(\left|0\right\rangle,-1)\otimes
O^B_1(\left|0\right\rangle,-1)$ and
$\left|1\right\rangle_B\leftrightarrow\left|2\right\rangle_B$,
$\left|1\right\rangle_C\leftrightarrow\left|2\right\rangle_C$. \\
(iii) $\left|\Omega_0\right\rangle_2=(a\left|0\right\rangle
+b\left|1\right\rangle)\left|33\right\rangle+\left|\varphi_2\right\rangle$.
Due to the invariance of $\left|\varphi_2\right\rangle$, choose
$b\neq0$. By using of the ILO's
$O^A_2(\left|1\right\rangle,-\alpha\left|0\right\rangle)\otimes
O^B_2(\left|0\right\rangle,\alpha\left|2\right\rangle)\otimes
O^C_2(\left|0\right\rangle,\alpha\left|2\right\rangle)$, we obtain
that $(\alpha\left|0\right\rangle
+\left|1\right\rangle)\left|33\right\rangle+\left|010\right\rangle+\left|001\right\rangle
+\left|112\right\rangle+\left|121\right\rangle\sim\left|\Phi_4\right\rangle$.\\
(iv) $\left|\Omega_0\right\rangle_3=(a\left|0\right\rangle
+b\left|1\right\rangle)\left|33\right\rangle+\left|\varphi_3\right\rangle$.
If $a=0$, then
$\left|\Omega_0\right\rangle_3\sim\left|\Phi_5\right\rangle$. For
the case of $a\neq0$, the operations
$O^A_2(\left|0\right\rangle,-\alpha\left|1\right\rangle)\otimes
O^B_2(\left|2\right\rangle,\alpha\left|0\right\rangle)\otimes
O^C_2(\left|2\right\rangle,\alpha\left|0\right\rangle)$ make
$(\left|0\right\rangle
+\alpha\left|1\right\rangle)\left|33\right\rangle+\left|100\right\rangle+\left|010\right\rangle+\left|001\right\rangle
+\left|112\right\rangle+\left|121\right\rangle\sim\left|\Phi_6\right\rangle$.\\
(v) $\left|\Omega_0\right\rangle_4=(a\left|0\right\rangle
+b\left|1\right\rangle)\left|33\right\rangle+\left|\varphi_4\right\rangle$.
If $b\neq0$, by the ILO's
$O^A_2(\left|1\right\rangle,-\alpha\left|0\right\rangle)\otimes
O^B_2(\left|1\right\rangle,\alpha\left|0\right\rangle)$, we obtain
$(\alpha\left|0\right\rangle
+\left|1\right\rangle)\left|33\right\rangle+\left|100\right\rangle+\left|010\right\rangle+\left|001\right\rangle
+\left|022\right\rangle\sim\left|\Phi_7\right\rangle$. The case of
$b=0$ is immediately taken into the form of
$\left|\Phi_8\right\rangle$.\\
(vi) $\left|\Omega_0\right\rangle_5=(a\left|0\right\rangle
+b\left|1\right\rangle)\left|33\right\rangle+\left|\varphi_5\right\rangle$.
If $b=0$, the ILO's
$\left|2\right\rangle_B\leftrightarrow\left|3\right\rangle_B$ and
$\left|2\right\rangle_C\leftrightarrow\left|3\right\rangle_C$
bring $\left|\Omega_0\right\rangle_5$ into the case of (v), while
the class $\left|\Phi_9\right\rangle$ is derived from the case of
$a=0$. Moreover in the case of $ab\neq0$, by performing the ILO's
$O^A_1(\left|0\right\rangle,\alpha^{-1})\otimes
O^B_1(\left|1\right\rangle,\alpha)\otimes
O^C_1(\left|1\right\rangle,\alpha)$ on
$\left|\Omega_0\right\rangle_5$, it leads to the class
$\left|\Phi_{10}\right\rangle$.

So far we have accomplished the computation of
$\left|\Omega_0\right\rangle$. Next, let us go on with the case of
$\left|\Omega_2\right\rangle$, which is more involved than the
above process. For $b\neq0$, we can move it away by the ILO's.
Hence,
\\
(i) $\left|\Omega_2\right\rangle_0=(\alpha\left|0\right\rangle
+\left|1\right\rangle)\left|33\right\rangle+\left|000\right\rangle+\left|111\right\rangle
+\left|022\right\rangle+\left|03\right\rangle\sum^2_{i=0}a_i\left|i\right\rangle$.
By the ILO's
$O^B_2(\left|0\right\rangle,-a_0\left|3\right\rangle)$,
$\left|0\right\rangle_B\leftrightarrow\left|3\right\rangle_B$ and
$\left|0\right\rangle_C\leftrightarrow\left|3\right\rangle_C$, one
can transform $\left|\Omega_2\right\rangle_0$ into the case of
$\left|\Omega_0\right\rangle$.\\
(ii) $\left|\Omega_2\right\rangle_1=(\alpha\left|0\right\rangle
+\left|1\right\rangle)\left|33\right\rangle+\left|000\right\rangle+\left|111\right\rangle
+(\left|0\right\rangle+\left|1\right\rangle)\left|22\right\rangle
+\left|03\right\rangle\sum^2_{i=0}a_i\left|i\right\rangle$. The
same ILO's make
$\left|\Omega_2\right\rangle_1\sim\left|\Omega_0\right\rangle$. \\
(iii) $\left|\Omega_2\right\rangle_2=(\alpha\left|0\right\rangle
+\left|1\right\rangle)\left|33\right\rangle+\left|010\right\rangle+\left|001\right\rangle
+\left|112\right\rangle+\left|121\right\rangle
+\left|03\right\rangle\sum^2_{i=0}a_i\left|i\right\rangle$. Due to
the operations
$O^A_2(\left|1\right\rangle,-\alpha\left|0\right\rangle)\otimes
O^B_2(\left|0\right\rangle,\alpha\left|2\right\rangle)\otimes
O^B_2(\left|0\right\rangle,\alpha\left|2\right\rangle)$, it holds
that $\left|\Omega_2\right\rangle_2\sim
\left|133\right\rangle+\left|010\right\rangle+\left|001\right\rangle
+\left|112\right\rangle+\left|121\right\rangle
+\left|03\right\rangle\sum^2_{i=0}a^{\prime}_i\left|i\right\rangle$,
which can be taken into the class $\left|\Phi_{11}\right\rangle$
by the ILO's
$O^B_2(\left|1\right\rangle,-a^{\prime}_0\left|3\right\rangle)\otimes
O^B_2(\left|0\right\rangle,-a^{\prime}_1\left|3\right\rangle)\otimes
O^C_2(\left|3\right\rangle,a^{\prime}_0\left|2\right\rangle)$.\\
(iv) $\left|\Omega_2\right\rangle_3=(\alpha\left|0\right\rangle
+\left|1\right\rangle)\left|33\right\rangle+\left|100\right\rangle+\left|010\right\rangle+\left|001\right\rangle
+\left|112\right\rangle+\left|121\right\rangle
+\left|03\right\rangle\sum^2_{i=0}a_i\left|i\right\rangle$. If
$\alpha=0$, the operations
$O^B_2(\left|1\right\rangle,-a_0\left|3\right\rangle)\otimes
O^B_2(\left|0\right\rangle,-a_1\left|3\right\rangle)\otimes
O^C_2(\left|3\right\rangle,a_0\left|2\right\rangle+a_1\left|0\right\rangle)$
make that
$\left|\Omega_2\right\rangle_3\sim\left|\Phi_{12}\right\rangle$.
For the case of $\alpha\neq0$, we can write out
$\left|\Omega_2\right\rangle_3\sim(\left|0\right\rangle
+\alpha\left|1\right\rangle)\left|33\right\rangle+\left|100\right\rangle+\left|010\right\rangle+\left|001\right\rangle
+\left|112\right\rangle+\left|121\right\rangle
+\left|03\right\rangle\sum^2_{i=0}a_i\left|i\right\rangle$, on
which we perform the ILO's
$O^A_2(\left|0\right\rangle,-\alpha\left|1\right\rangle)\otimes
O^C_2(\left|3\right\rangle,-\sum^2_{i=0}a_i\left|i\right\rangle)\otimes
O^B_2(\left|2\right\rangle,\alpha\left|0\right\rangle)\otimes
O^C_2(\left|2\right\rangle,\alpha\left|0\right\rangle)$ and get
$\left|\Omega_2\right\rangle_3\sim\left|033\right\rangle+\left|100\right\rangle+\left|010\right\rangle
+\left|001\right\rangle+\left|112\right\rangle+\left|121\right\rangle
+\left|13\right\rangle\sum^2_{i=0}a^{\prime}_i\left|i\right\rangle$.
Fortunately, this expression can be brought into the case of
$\left|\Omega_0\right\rangle$ by the ILO's
$O^B_2(\left|0\right\rangle,-a^{\prime}_0\left|3\right\rangle)\otimes
O^B_2(\left|1\right\rangle,-a^{\prime}_2\left|3\right\rangle)\otimes
O^B_2(\left|2\right\rangle,-a^{\prime}_1\left|3\right\rangle)\otimes
O^C_2(\left|3\right\rangle,a^{\prime}_2\left|0\right\rangle+a^{\prime}_0\left|1\right\rangle)$.\\
(v) $\left|\Omega_2\right\rangle_4=(\alpha\left|0\right\rangle
+\left|1\right\rangle)\left|33\right\rangle+\left|100\right\rangle+\left|010\right\rangle+\left|001\right\rangle
+\left|022\right\rangle+\left|03\right\rangle\sum^2_{i=0}a_i\left|i\right\rangle$.
Similar to (i) and (ii), performing the operations
$O^B_2(\left|2\right\rangle,-a_2\left|3\right\rangle)$,
$\left|2\right\rangle_B\leftrightarrow\left|3\right\rangle_B$ and
$\left|2\right\rangle_C\leftrightarrow\left|3\right\rangle_C$
makes that
$\left|\Omega_2\right\rangle_4\sim\left|\Omega_0\right\rangle$.\\
(vi) $\left|\Omega_2\right\rangle_5=(\alpha\left|0\right\rangle
+\left|1\right\rangle)\left|33\right\rangle+\left|100\right\rangle+\left|010\right\rangle+\left|001\right\rangle
+\left|122\right\rangle+\left|03\right\rangle\sum^2_{i=0}a_i\left|i\right\rangle$.
If $\alpha=0$, by the ILO's
$O^B_2(\left|1\right\rangle,-a_0\left|3\right\rangle)\otimes
O^B_2(\left|0\right\rangle,-a_1\left|3\right\rangle)\otimes
O^C_2(\left|3\right\rangle,a_1\left|0\right\rangle)$, we obtain
$\left|\Omega_2\right\rangle_5\sim\left|\Phi_{14}\right\rangle$.
For the case of $\alpha\neq0$, we first perform the operations
$O^A_2(\left|1\right\rangle,-\alpha\left|0\right\rangle)\otimes
O^C_2(\left|1\right\rangle,\alpha\left|0\right\rangle)$ so that
$\left|\Omega_2\right\rangle_5=\left|133\right\rangle+\left|100\right\rangle+\left|010\right\rangle
+\left|001\right\rangle+\left|122\right\rangle-\alpha\left|022\right\rangle
+\left|03\right\rangle\sum^2_{i=0}a^{\prime}_i\left|i\right\rangle$.
Subsequently, by the ILO's
$O^B_2(\left|2\right\rangle,a^{\prime}_2/\alpha\left|3\right\rangle)\otimes
O^C_2(\left|3\right\rangle,-a^{\prime}_2/\alpha\left|2\right\rangle)$,
$\left|2\right\rangle_B\leftrightarrow\left|3\right\rangle_B$ and
$\left|2\right\rangle_C\leftrightarrow\left|3\right\rangle_C$, we
transform the above situation to that of
$\left|\Omega_0\right\rangle$.

With the help of the results of both $\left|\Omega_0\right\rangle$
and $\left|\Omega_2\right\rangle$, we can treat the case of
$\left|\Omega_3\right\rangle$ more succinctly. In the same way,
let $\left|\Omega_3\right\rangle_i=\left|\Omega_0\right\rangle_i
+\left|13\right\rangle\left|\chi\right\rangle.$ For the case of
$\left|\Omega_3\right\rangle_i,i=0,5$, the tricks similar to that
of $\left|\Omega_2\right\rangle_0$ and
$\left|\Omega_2\right\rangle_1$ realize that
$\left|\Omega_3\right\rangle_0\sim\left|\Omega_0\right\rangle$,
while the invariance of $\left|\varphi_1\right\rangle$ and
$\left|\varphi_2\right\rangle$ has implied
$\left|\Omega_3\right\rangle_i\sim\left|\Omega_2\right\rangle_i,i=1,2.$
Subsequently, $\left|\Omega_3\right\rangle_3$ can be taken into
the form of $\left|\Omega_0\right\rangle$ by the skills similar to
that of the case of $\alpha\neq0$ of
$\left|\Omega_2\right\rangle_3$. The sole new class is derived
from the calculation of
$\left|\Omega_3\right\rangle_4=(\left|0\right\rangle+\alpha\left|1\right\rangle)\left|33\right\rangle
+\left|100\right\rangle+\left|010\right\rangle+\left|001\right\rangle
+\left|022\right\rangle+\left|13\right\rangle\sum^2_{i=0}a_i\left|i\right\rangle$,
which can be taken into the form of
$\left|\Omega_2\right\rangle_4$ if $\alpha\neq0$, by the ILO's
$O^A_2(\left|1\right\rangle,-1/\alpha\left|0\right\rangle)\otimes
O^B_2(\left|1\right\rangle,1/\alpha\left|0\right\rangle)\otimes
O^C_2(\left|3\right\rangle,-1/\alpha\sum^2_{i=0}a_i\left|i\right\rangle)$.
On the other hand let $\alpha=0$. If $a_1\neq0$, in virtue of the
operations
$O^C_2(\left|1\right\rangle,-a_2/a_1\left|2\right\rangle)\otimes
O^B_2(\left|2\right\rangle,a_2/a_1\left|0\right\rangle)$,
$\left|2\right\rangle_B\leftrightarrow\left|3\right\rangle_B$ and
$\left|2\right\rangle_C\leftrightarrow\left|3\right\rangle_C$ we
make that
$\left|\Omega_3\right\rangle_4\sim\left|\Omega_0\right\rangle$.
For the case of $a_1=0$, performing the operations
$O^B_2(\left|0\right\rangle,-a_0\left|3\right\rangle)\otimes
O^C_2(\left|2\right\rangle,a_0\left|1\right\rangle)$ on
$\left|\Omega_3\right\rangle_4$ gives rise to the fact that
$\left|\Omega_3\right\rangle_4\sim\left|\Phi_{13}\right\rangle$.
Finally we calculation the family of
$\left|\Omega_1\right\rangle$, which only contains two subcases.
For the case of
$\left|\Omega_1\right\rangle_0=\left|033\right\rangle+\left|132\right\rangle+\left|\phi_0\right\rangle$,
we can transform it into the family of
$\left|\Omega_0\right\rangle$ by the operations
$\left|0\right\rangle_B\leftrightarrow\left|3\right\rangle_B$ and
$\left|0\right\rangle_C\leftrightarrow\left|3\right\rangle_C$. The
case of
$\left|\Omega_1\right\rangle_1=\left|033\right\rangle+\left|132\right\rangle+\left|\phi_1\right\rangle$
is just the class $\left|\Phi_{15}\right\rangle$.

In what follows, the task is to prove that these 16 entangled
classes are essentially inequivalent under SLOCC. In principle, we
distinguish the classes by range criterion, and specially in terms
of the notation $[\ast,\ast,\ast]$ defined in range criterion. As
the situation is more complicated than those in \cite{Chen}, we
will fully employ the fact that the local rank of subsystem is
invariant under ILO's. For example, consider the two classes
\begin{eqnarray}
\left|\Phi_0\right\rangle&=&\left|133\right\rangle+
\left|000\right\rangle+\left|111\right\rangle
+\left|022\right\rangle,\nonumber\\
\left|\Phi_2\right\rangle&=&\left|133\right\rangle+
\left|000\right\rangle+\left|111\right\rangle
+(\left|0\right\rangle+\left|1\right\rangle)\left|22\right\rangle,
\end{eqnarray}
which can be rewritten as
\begin{eqnarray}
\left|\Phi_0\right\rangle_{ABC}&=&\left|0\right\rangle(\left|00\right\rangle+\left|22\right\rangle)
+\left|1\right\rangle(\left|11\right\rangle+\left|33\right\rangle),\nonumber\\
\left|\Phi_2\right\rangle_{ABC}&=&\left|0\right\rangle(\left|00\right\rangle+\left|22\right\rangle)
+\left|1\right\rangle(\left|11\right\rangle+\left|22\right\rangle+\left|33\right\rangle).\nonumber\\
\end{eqnarray}
Evidently, the local rank of the state in $R(\rho^{BC}_{\Phi_0})$
is 2 or 4, while that in $R(\rho^{BC}_{\Phi_1})$ can be 2, 3 or 4.
Thus the two states are inequivalent. For simplicity we define the
set $(a_0,a_1,...)$, and a state $\left|\Psi\right\rangle_{ABC}$
belongs to this set iff all the possible values of local ranks of
the states in $R(\rho^{BC}_{\Psi})$ are $a_0,a_1,...$. Combined
with the notation $[\ast,\ast,\ast]$ we have
\begin{eqnarray}
\left|\Phi_0\right\rangle_{ABC}&\in&[0,\infty,\infty]\cap(2,4),\nonumber\\
\left|\Phi_2\right\rangle_{ABC}&\in&[0,\infty,\infty]\cap(2,3,4).
\end{eqnarray}

\begin{center}
{\bf A. Discrimination of the classes in $[0,\infty,\infty]$,
$[1,\infty,\infty]$ and $[0,2,2]$}
\end{center}

There are nine entangled classes in the set of
$[0,\infty,\infty]$. We list them with the notation of local ranks
$(a_0,a_1,...)$,
\begin{eqnarray}
\left|\Phi_0\right\rangle&=&\left|133\right\rangle+
\left|\varphi_0\right\rangle\in(2,4);\nonumber\\
\left|\Phi_2\right\rangle&=&\left|133\right\rangle+
\left|\varphi_1\right\rangle\in(2,3,4);\nonumber\\
\left|\Phi_4\right\rangle&=&\left|133\right\rangle+
\left|\varphi_2\right\rangle\in(2,3);\nonumber\\
\left|\Phi_5\right\rangle&=&\left|133\right\rangle+
\left|\varphi_3\right\rangle\in(2,4);\nonumber\\
\left|\Phi_7\right\rangle&=&\left|133\right\rangle+
\left|\varphi_4\right\rangle\in(2,3,4);\nonumber\\
\left|\Phi_9\right\rangle&=&\left|133\right\rangle+
\left|\varphi_5\right\rangle\in(2,3,4);\nonumber\\
\left|\Phi_{11}\right\rangle&=&\left|133\right\rangle+\left|032\right\rangle+
\left|\varphi_2\right\rangle\in(3);\nonumber\\
\left|\Phi_{13}\right\rangle&=&\left|033\right\rangle+\left|132\right\rangle+
\left|\varphi_4\right\rangle\in(2,4);\nonumber\\
\left|\Phi_{15}\right\rangle&=&\left|033\right\rangle+\left|132\right\rangle+
\left|\phi_1\right\rangle\in(3).
\end{eqnarray}
Except the sole class $\left|\Phi_4\right\rangle$ with (2,3), the
other eight classes can be divided into three groups with respect
to the local ranks, i.e., (2,4), (2,3,4) and (3).

I. the case of (2,4), including three classes
$\left|\Phi_0\right\rangle$, $\left|\Phi_5\right\rangle$ and
$\left|\Phi_{13}\right\rangle$. Notice that there are two rank-2
states $\left|00\right\rangle+\left|22\right\rangle$ and
$\left|11\right\rangle+\left|33\right\rangle$ in
$R(\rho^{BC}_{\Phi_0})$, while there is only one such state in
$R(\rho^{BC}_{\Phi_5})$ and $R(\rho^{BC}_{\Phi_{13}})$
respectively. Due to the range criterion,
$\left|\Phi_0\right\rangle$ differs from the other two states. In
order to compare $\left|\Phi_5\right\rangle$ and
$\left|\Phi_{13}\right\rangle$, write out
\begin{eqnarray}
R(\rho^{AB}_{\Phi_5})&=&a\left|13\right\rangle+b\left|11\right\rangle
+c(\left|00\right\rangle+\left|12\right\rangle)+d(\left|10\right\rangle+\left|01\right\rangle)
,\nonumber\\
R(\rho^{AB}_{\Phi_{13}})&=&a^{\prime}\left|03\right\rangle+b^{\prime}\left|00\right\rangle
+c^{\prime}(\left|13\right\rangle+\left|02\right\rangle)+d^{\prime}(\left|10\right\rangle+\left|01\right\rangle)
.\nonumber\\
\end{eqnarray}
Observe these two expressions. Although there are infinitely many
product states in either of them, it occurs iff $c=d=0$ and
$c^{\prime}=d^{\prime}=0$. This implies that the possible ILO's
must transform
$(a^{\prime}\left|3\right\rangle+b^{\prime}\left|0\right\rangle)^B_{\Phi_{13}}$
into $(a\left|3\right\rangle+b\left|1\right\rangle)^B_{\Phi_5}$,
which means either $\left|3\right\rangle^B_{\Phi_{13}}$ or
$\left|0\right\rangle^B_{\Phi_{13}}$ will be taken into
$(\left|3\right\rangle+\alpha\left|1\right\rangle)^B_{\Phi_5}$.
However, the adjoint state of $\left|3\right\rangle^B_{\Phi_{13}}$
is $\left|03\right\rangle+\left|12\right\rangle$ and that of
$\left|0\right\rangle^B_{\Phi_{13}}$ is
$\left|10\right\rangle+\left|01\right\rangle$, and either of these
two adjoint states makes the range an entangled state. Since the
adjoint state of $\left|3\right\rangle^B_{\Phi_5}$ is of product
form, we get the outcome that $\left|\Phi_5\right\rangle$ and
$\left|\Phi_{13}\right\rangle$ are not equivalent under SLOCC.

II. the case of (2,3,4), including three classes
$\left|\Phi_2\right\rangle$, $\left|\Phi_7\right\rangle$ and
$\left|\Phi_9\right\rangle$. Similar to the case of I,
$\left|\Phi_2\right\rangle$ is a distinctive class containing two
rank-3 states in $R(\rho^{BC}_{\Phi_2})$, while there is only one
such state in $R(\rho^{BC}_{\Phi_7})$ and $R(\rho^{BC}_{\Phi_9})$
respectively. In order to compare $\left|\Phi_7\right\rangle$ and
$\left|\Phi_9\right\rangle$, write out
\begin{eqnarray}
R(\rho^{AB}_{\Phi_7})&=&a\left|13\right\rangle+b\left|02\right\rangle
+c\left|00\right\rangle+d(\left|10\right\rangle+\left|01\right\rangle)
,\nonumber\\
R(\rho^{AB}_{\Phi_9})&=&a^{\prime}\left|13\right\rangle+b^{\prime}\left|12\right\rangle
+c^{\prime}\left|00\right\rangle+d^{\prime}(\left|10\right\rangle+\left|01\right\rangle)
.\nonumber\\
\end{eqnarray}
Notice the possible ILO's must make that
$\left|0\right\rangle^A_{\Phi_7}\leftrightarrow\left|1\right\rangle^A_{\Phi_9}$
and
$\left|1\right\rangle^A_{\Phi_7}\leftrightarrow\left|0\right\rangle^A_{\Phi_9}$
due to the local ranks of the adjoint states. By the expressions
of $R(\rho^{AB}_{\Phi_7})$ and $R(\rho^{AB}_{\Phi_9})$ we have
$\left|0\right\rangle^B_{\Phi_9}\rightarrow\left|3\right\rangle^B_{\Phi_7}$.
Since the adjoint state of $\left|0\right\rangle^B_{\Phi_9}$ is
$\left|10\right\rangle+\left|01\right\rangle$ making the range
entangled while the adjoint state of
$\left|3\right\rangle^B_{\Phi_7}$ is of product form, the two
states $\left|\Phi_7\right\rangle$ and $\left|\Phi_9\right\rangle$
are inequivalent.

III. the case of (3), including two classes
$\left|\Phi_{11}\right\rangle$ and $\left|\Phi_{15}\right\rangle$.
Write out
\begin{eqnarray}
R(\rho^{AB}_{\Phi_{11}})&=&a\left|13\right\rangle+b\left|01\right\rangle
+c(\left|03\right\rangle+\left|11\right\rangle)+d(\left|00\right\rangle+\left|12\right\rangle)
,\nonumber\\
R(\rho^{AB}_{\Phi_{15}})&=&a^{\prime}\left|03\right\rangle+b^{\prime}\left|13\right\rangle
+c^{\prime}(\left|00\right\rangle+\left|11\right\rangle)+d^{\prime}(\left|01\right\rangle+\left|12\right\rangle)
.\nonumber\\
\end{eqnarray}
Evidently, the product states in $R(\rho^{AB}_{\Phi_{13}})$ are
$a^{\prime}\left|03\right\rangle+b^{\prime}\left|13\right\rangle$,
while there exist the product states $\left|13\right\rangle$ and
$\left|01\right\rangle$ in $R(\rho^{AB}_{\Phi_{11}})$, and there
does not exist an ILO making $\left|3\right\rangle^B_{\Phi_{15}}$
to $\left|3\right\rangle^B_{\Phi_{11}}$ and
$\left|1\right\rangle^B_{\Phi_{11}}$ simultaneously. Consequently,
$\left|\Phi_{11}\right\rangle$ and $\left|\Phi_{15}\right\rangle$
differs from each other.

Therefore we have accomplished the task of comparison of the
entangled classes in $[0,\infty,\infty]$. With the similar
techniques we can easily treat the residual business. There are
two classes belonging to $[1,\infty,\infty]$ as follows,
\begin{eqnarray}
\left|\Phi_1\right\rangle_{ABC}&=&\left|033\right\rangle
+\left|000\right\rangle+\left|111\right\rangle+\left|022\right\rangle\in(1,3,4),\nonumber\\
\left|\Phi_8\right\rangle_{ABC}&=&\left|033\right\rangle
+\left|100\right\rangle+\left|010\right\rangle
+\left|001\right\rangle+\left|022\right\rangle\in(1,4),\nonumber\\
\end{eqnarray}
and one can immediately recognize the inequivalence of them. For
the case of $[0,2,2]$, we also write the concrete forms of two
classes both of which belong to $(3,4)$
\begin{eqnarray}
\left|\Phi_6\right\rangle_{ABC}&=&\left|033\right\rangle
+\left|100\right\rangle+\left|010\right\rangle
+\left|001\right\rangle+\left|112\right\rangle+\left|121\right\rangle,\nonumber\\
\left|\Phi_{14}\right\rangle_{ABC}&=&\left|133\right\rangle
+\left|032\right\rangle+\left|100\right\rangle+\left|010\right\rangle
+\left|001\right\rangle+\left|122\right\rangle.\nonumber\\
\end{eqnarray}
It is easy to find out the two product states in
$R(\rho^{AC}_{\Phi_6})$ are $\left|03\right\rangle$ and
$\left|11\right\rangle$, and that in $R(\rho^{AC}_{\Phi_{14}})$
are $\left|00\right\rangle$ and $\left|12\right\rangle$. Thus the
possible ILO's must bring $\left|0\right\rangle^C_{\Phi_{14}}$ or
$\left|2\right\rangle^C_{\Phi_{14}}$ into
$\left|3\right\rangle^C_{\Phi_6}$. Since either of the two adjoint
states of $\left|0\right\rangle^C_{\Phi_{14}}$ and
$\left|2\right\rangle^C_{\Phi_{14}}$ makes the range entangled,
while the adjoint state of $\left|3\right\rangle^C_{\Phi_6}$ is of
product form, so $\left|\Phi_6\right\rangle$ and
$\left|\Phi_{14}\right\rangle$ are inequivalent. As the other
three classes $\left|\Phi_3\right\rangle_x$,
$\left|\Phi_{10}\right\rangle$ and $\left|\Phi_{12}\right\rangle$
belong to different sets, we thus assert that there are indeed 16
classes of states in true $2\times4\times4$ space, under the SLOCC
criterion.

\begin{center}
{\bf B. Classification of the entangled class
$\left|\Phi_3\right\rangle$}
\end{center}

In this subsection, we investigate the structure of the class
$\left|\Phi_3\right\rangle_x=(\left|0\right\rangle+x\left|1\right\rangle)\left|33\right\rangle+
\left|000\right\rangle+\left|111\right\rangle+
(\left|0\right\rangle+\left|1\right\rangle)\left|22\right\rangle
$, which contains a parameter $x\neq0,1$. Evidently, the set of
product states in $R(\rho^{AB}_{\Phi_{3x}})$ is $S_x=
\{(\left|0\right\rangle+x\left|1\right\rangle)\left|3\right\rangle,
\left|00\right\rangle,\left|11\right\rangle
,(\left|0\right\rangle+\left|1\right\rangle)\left|2\right\rangle\}$.
Thus, if
$\left|\Phi_3\right\rangle_x\sim\left|\Phi_3\right\rangle_y$, then
the possible ILO's must bring $S_x$ into $S_y$, i.e., every
element in $S_x$ will be transformed into some element in $S_y$.
Due to the symmetry of system $BC$, it holds that every term in
$\left|\Phi_3\right\rangle_x$ will be transformed into some term
in $\left|\Phi_3\right\rangle_y$, e.g.,
$(\left|0\right\rangle+x\left|1\right\rangle)\left|33\right\rangle
\rightarrow(\left|0\right\rangle+y\left|1\right\rangle)\left|33\right\rangle,$
$\left|000\right\rangle_x\rightarrow\left|000\right\rangle_y,$
$\left|111\right\rangle_x\rightarrow\left|111\right\rangle_y,$
$(\left|0\right\rangle+\left|1\right\rangle)\left|2\right\rangle_x
\rightarrow(\left|0\right\rangle+\left|1\right\rangle)\left|2\right\rangle_y$.
It then seems that we have to treat 24 subcases with respect to
all kinds of matches, but actually only three of them is worth a
further investigation. This can be seen as follows,
\begin{eqnarray}
V_{ABC}[\underbrace{(\left|0\right\rangle+x\left|1\right\rangle)\left|33\right\rangle}_{I}+
\underbrace{\left|000\right\rangle}_{II}+\underbrace{\left|111\right\rangle}_{III}+
\underbrace{(\left|0\right\rangle+\left|1\right\rangle)\left|22\right\rangle}_{IV}]\nonumber\\
=\underbrace{(\left|0\right\rangle+y\left|1\right\rangle)\left|33\right\rangle}_{1}+
\underbrace{\left|000\right\rangle}_{2}+\underbrace{\left|111\right\rangle}_{3}+
\underbrace{(\left|0\right\rangle+\left|1\right\rangle)\left|22\right\rangle}_{4},\nonumber\\
\end{eqnarray}
where $V_{ABC}=V_A\otimes V_B\otimes V_C$. Another alternative
expression of this equation is
\begin{eqnarray}
V^{\prime}_{ABC}[\underbrace{(\left|0\right\rangle+x^{-1}\left|1\right\rangle)\left|33\right\rangle}_{I}+
\underbrace{\left|111\right\rangle}_{II}+\underbrace{\left|000\right\rangle}_{III}+
\underbrace{(\left|0\right\rangle+\left|1\right\rangle)\left|22\right\rangle}_{IV}]\nonumber\\
=\underbrace{(\left|0\right\rangle+y^{-1}\left|1\right\rangle)\left|33\right\rangle}_{1}+
\underbrace{\left|111\right\rangle}_{2}+\underbrace{\left|000\right\rangle}_{3}+
\underbrace{(\left|0\right\rangle+\left|1\right\rangle)\left|22\right\rangle}_{4},\nonumber\\
\end{eqnarray}
where
$V^{\prime}_{ABC}=\sigma_{xA}\otimes\sigma_{xB}\otimes(O^C_1(\left|3\right\rangle,y^{-1})\sigma_{xC})(V_A\otimes
V_B\otimes
V_C)\sigma_{xA}\otimes\sigma_{xB}\otimes(O^C_1(\left|3\right\rangle,x)\sigma_{xC})$
and $\sigma_x$ is the Pauli operator. Similarly one can choose
$V^{\prime\prime}_{ABC}=(V_A\otimes V_B\otimes
V_C)\sigma_{xA}\otimes\sigma_{xB}\otimes(O^C_1(\left|3\right\rangle,x)\sigma_{xC})$
or
$V^{\prime\prime\prime}_{ABC}=\sigma_{xA}\otimes\sigma_{xB}\otimes(O^C_1(\left|3\right\rangle,y^{-1})\sigma_{xC})
(V_A\otimes V_B\otimes V_C)$. We regard the position of every term
invariant under either of these four transformations, e.g.,
$I\rightarrow2,II\rightarrow3,III\rightarrow4,IV\rightarrow1,$
although the contents of them are changed. We can bring this match
into the case of
$\left|000\right\rangle_x\rightarrow\left|000\right\rangle_{y^{-1}}$
and
$\left|111\right\rangle_x\rightarrow(\left|0\right\rangle+\left|1\right\rangle)\left|22\right\rangle_{y^{-1}}$
by $V^{\prime\prime\prime}_{ABC}$. In the same vein, by virtue of
these four transformations and the invariance of the matches, all
the 24 subcases can be taken into three scenarios: (i)
$II,III\rightarrow2,3$; (ii) $II\rightarrow2,3$ and
$III\rightarrow1,4$; (iii) $II,III\rightarrow1,4$, where we only
need to replace the parameter $x$ in the final result with
$x^{-1}$ as well as the parameter $y$ with $y^{-1}$, e.g., the
equation $x=y$ contains other three equations $x^{-1}=y$,
$x=y^{-1}$ and $x^{-1}=y^{-1}$, so that all possible value of $y$
can be achieved. For the case of (i), we can get the following
results after some ILO's
\begin{eqnarray}
\left|\Phi_3\right\rangle^{0}_x&=&(a\left|0\right\rangle+x\left|1\right\rangle)\left|33\right\rangle+
\left|000\right\rangle+\left|111\right\rangle+
(a\left|0\right\rangle+\left|1\right\rangle)\left|22\right\rangle,\nonumber\\
\left|\Phi_3\right\rangle^{1}_x&=&(bx\left|0\right\rangle+\left|1\right\rangle)\left|33\right\rangle+
\left|000\right\rangle+\left|111\right\rangle+
(b\left|0\right\rangle+\left|1\right\rangle)\left|22\right\rangle,\nonumber\\
\end{eqnarray}
where some coefficients have been moved away by the ILO's $O_1$
and $a,b$ are the residual parameters. Compare these two
expressions with that of $\left|\Phi_3\right\rangle_y$
supplemented by the possible ILO's
$\left|2\right\rangle_B\leftrightarrow\left|3\right\rangle_B$ and
$\left|2\right\rangle_C\leftrightarrow\left|3\right\rangle_C$, we
can get $y=x,x^{-1}$. For the case of (ii), it is easy to choose a
definite transformation
$\left|000\right\rangle_{x^{\prime}}\rightarrow\left|000\right\rangle_{y^{\prime}}$
and $\left|111\right\rangle_{x^{\prime}}\rightarrow
(\left|0\right\rangle+\left|1\right\rangle)\left|22\right\rangle_{y^{\prime}}$,
where $x^{\prime}=x,x^{-1}$ and $y^{\prime}=y,y^{-1}$.
Consequently, the resulting state is
\begin{eqnarray}
\left|\Phi_3\right\rangle_{y^{\prime}}&=&(a\left|0\right\rangle
+x^{\prime}(\left|0\right\rangle+\left|1\right\rangle))\left|33\right\rangle+\left|000\right\rangle\nonumber\\
&+&(\left|0\right\rangle+\left|1\right\rangle)\left|22\right\rangle+
(a\left|0\right\rangle+(\left|0\right\rangle+\left|1\right\rangle))\left|11\right\rangle,\nonumber\\
\end{eqnarray}
If $a=-1$ we have
\begin{eqnarray}
\left|\Phi_3\right\rangle_{y^{\prime}}&\sim&
(\frac{x^{\prime}-1}{x^{\prime}}\left|0\right\rangle+\left|1\right\rangle)\left|33\right\rangle
+\left|000\right\rangle\nonumber\\
&+&(\left|0\right\rangle+\left|1\right\rangle)\left|22\right\rangle+
\left|111\right\rangle,
\end{eqnarray}
and hence $\frac{x^{\prime}-1}{x^{\prime}}=y^{\prime}$, or
$y=1-x,1-x^{-1},(1-x)^{-1},(1-x^{-1})^{-1}.$ For the case of
$a=-x^{\prime}$, do the exchange between
$\left|11\right\rangle_{BC}$ and $\left|33\right\rangle_{BC}$ and
compare the resulting expression with that of
$\left|\Phi_3\right\rangle_{y^{\prime}}$, we get
$1-x^{\prime}=y^{\prime}$ which leads to the same result above.
While in the case of (iii), choose
$\left|000\right\rangle_{x^{\prime}}\rightarrow
(\left|0\right\rangle+y^{\prime}\left|1\right\rangle)\left|33\right\rangle_{y^{\prime}}$
and $\left|111\right\rangle_{x^{\prime}}\rightarrow
(\left|0\right\rangle+\left|1\right\rangle)\left|22\right\rangle_{y^{\prime}}$,
and the resulting state is
\begin{eqnarray}
\left|\Phi_3\right\rangle_{y^{\prime}}&=&(a(\left|0\right\rangle+y^{\prime}\left|1\right\rangle)
+x^{\prime}(\left|0\right\rangle+\left|1\right\rangle))\left|00\right\rangle\nonumber\\
&+&(\left|0\right\rangle+y^{\prime}\left|1\right\rangle)\left|33\right\rangle
+(\left|0\right\rangle+\left|1\right\rangle)\left|22\right\rangle\nonumber\\
&+&(a(\left|0\right\rangle+y^{\prime}\left|1\right\rangle)
+(\left|0\right\rangle+\left|1\right\rangle))\left|11\right\rangle.
\end{eqnarray}
Follow the same technique in (ii), choose
$a=-1,-{y^{\prime}}^{-1}$ such that
$y^{\prime}=x^{\prime},{x^{\prime}}^{-1}$, which again leads to
$y=x,x^{-1}$. In all, we find that
$\left|\Phi_3\right\rangle_x\sim\left|\Phi_3\right\rangle_{1-x}\sim\left|\Phi_3\right\rangle_{1-x^{-1}}\sim
\left|\Phi_3\right\rangle_{x^{-1}}\sim\left|\Phi_3\right\rangle_{(1-x)^{-1}}
\sim\left|\Phi_3\right\rangle_{(1-x^{-1})^{-1}}.$ Define the set
$C_x\equiv\{x,x^{-1},1-x,(1-x)^{-1},1-x^{-1},(1-x^{-1})^{-1}\},x\neq0,1$,
then $\left|\Phi_3\right\rangle_x\sim\left|\Phi_3\right\rangle_y$
iff $y\in C_x$. It is easy to verify that the whole complex number
field can be expressed as $C=\{0,1\}\cup C_{x_0}\cup
C_{x_1}\cup\cdots,x_i\neq x_j$ for different $i,j.$ In addition,
any two elements chosen from distinct sets $C_{x_i}$ and $C_{x_j}$
are not identical. These characters state that the entangled class
$\left|\Phi_3\right\rangle_x$ can be divided into infinitely many
kinds of states.      Q.E.D.

\section{hierarchy of entanglement in $2\times3\times N$ space
and classification of $2\times(M+3)\times(2M+3)$ and
$2\times(M+4)\times(2M+4)$ states}

In the preceding section, we have proved that there exist
infinitely many entangled classes in the $2\times4\times4$ space
under SLOCC, for the existence of state
$\left|\Phi_3\right\rangle_x$. It is known that there is only a
finite number of essential classes in the $2\times 2\times
N,N=2,3,4,...$ and $2\times 3\times N,N=3,4,5,6...$ space
\cite{Dur,Miyake2,Chen}, so the $2\times4\times4$ states are the
simplest family containing an infinite number of entangled
classes. For some trivial cases, i.e., the unentangled states and
those product in one party and entangled with respect to the other
two, we omit the deduction of classification. Subsequently, we
generally list the classification of $2\times 3\times N$ states in
the following table.
\begin{tabular}{l|l}
true rank of system & Class\\
\hline $2\times3\times N (N\geq6)$ &
$\left|000\right\rangle+\left|011\right\rangle+\left|022\right\rangle+\left|103\right\rangle$\\
&$+\left|114\right\rangle+\left|125\right\rangle;$ \\
\hline$2\times3\times 5$ &
$\left|024\right\rangle+\left|000\right\rangle+\left|011\right\rangle$\\
&$+\left|102\right\rangle+\left|113\right\rangle;$\\
&$\left|024\right\rangle+\left|121\right\rangle+\left|000\right\rangle+\left|011\right\rangle$\\
&$+\left|102\right\rangle+\left|113\right\rangle;$\\
\end{tabular}

\begin{tabular}{l|l}
\hline$2\times3\times 4$ &
$\left|123\right\rangle+\left|012\right\rangle+\left|000\right\rangle+\left|101\right\rangle;$\\
& $\left|023\right\rangle+\left|012\right\rangle+\left|000\right\rangle+\left|101\right\rangle;$\\
&
$\left|123\right\rangle+\left|012\right\rangle+\left|110\right\rangle
+\left|000\right\rangle+\left|101\right\rangle;$\\
& $\left|023\right\rangle+\left|122\right\rangle
+\left|012\right\rangle+\left|000\right\rangle+\left|101\right\rangle;$\\
& $\left|023\right\rangle+\left|122\right\rangle
+\left|012\right\rangle+\left|110\right\rangle+\left|000\right\rangle+\left|101\right\rangle;$\\
\hline$2\times3\times 3$ &
$\left|000\right\rangle+\left|111\right\rangle+
\left|022\right\rangle;$\\
& $\left|000\right\rangle+\left|111\right\rangle+
\left|022\right\rangle+\left|122\right\rangle;$\\
& $\left|010\right\rangle+\left|001\right\rangle+
\left|112\right\rangle+\left|121\right\rangle;$\\
&$\left|100\right\rangle+\left|010\right\rangle+\left|001\right\rangle+
\left|112\right\rangle+\left|121\right\rangle;$\\
& $\left|100\right\rangle+\left|010\right\rangle+
\left|001\right\rangle+\left|022\right\rangle;$\\
& $\left|100\right\rangle+\left|010\right\rangle+
\left|001\right\rangle+\left|122\right\rangle;$\\
\hline $2\times3\times 2$ &
$\left|000\right\rangle+\left|011\right\rangle+\left|121\right\rangle;$\\
&$\left|000\right\rangle+\left|011\right\rangle+\left|110\right\rangle+\left|121\right\rangle;$\\
\hline $2\times2\times 4$ &
$\left|000\right\rangle+\left|011\right\rangle+\left|102\right\rangle+\left|113\right\rangle;$\\
\hline $2\times2\times 3$ &
$\left|000\right\rangle+\left|011\right\rangle+\left|112\right\rangle;$\\
&$\left|000\right\rangle+\left|011\right\rangle+\left|101\right\rangle+\left|112\right\rangle;$\\
\hline $2\times2\times 2$ &
$\left|000\right\rangle+\left|111\right\rangle;$\\
&$\left|001\right\rangle+\left|010\right\rangle+\left|100\right\rangle;$\\
\hline $1\times3\times 3$ &
$\left|000\right\rangle+\left|011\right\rangle+\left|022\right\rangle;$\\
\hline $1\times2\times 2$ &
$\left|000\right\rangle+\left|011\right\rangle;$\\
\hline $2\times1\times 2$ &
$\left|000\right\rangle+\left|101\right\rangle;$\\
\hline $2\times2\times 1$ &
$\left|000\right\rangle+\left|110\right\rangle;$\\
\hline $1\times1\times 1$ & $\left|000\right\rangle.$
\end{tabular}

Let us analyze the above table. By this hierarchy, there are
totally 26 entangled classes in the whole $2\times3\times N$
system under SLOCC. As mentioned in the LHRGM, we see that the
classes are generating in a regular way, i.e., higher entangled
classes are intimately concerned with lower entangled ones. For
example, the classes in $2\times3\times5$ space have a ``branch"
structure such that
\[
\overbrace{\left|000\right\rangle+\left|011\right\rangle
+\left|102\right\rangle+\left|113\right\rangle}^{2\times2\times4}
+\left\{\begin{array}{l}
\left|024\right\rangle,\\
\left|024\right\rangle+\left|121\right\rangle.
\end{array}
\right.
\]
Here, both of the states consist of two parts, including one
original $2\times2\times4$ class and the other portion by the
added dimension. According to the LHRGM, every product term of the
added portion always contains at least a new high ket, such as
$\left|2\right\rangle_B$, or $\left|4\right\rangle_C$ in the term
$\left|024\right\rangle$. This suggests that the added portion
could be viewed as a $distribution$ of all high kets, complemented
by the lower kets existing in the original class. By simply
writing all distributions, one can primarily realize the structure
of a certain kind of entanglement. On the other hand, the number
of inequivalent classes differs with the changing dimensions.
Specifically, there are orderly $1,2,6,5,2,1$ kinds of states in
$2\times3\times N$ space, $N=1,2,3,4,5,6$, which implies the
$2\times3\times3$ states has the most classes of entanglement. One
can understand this fact like this. As the A,B system is in the
$2\times3$ space, in which there are generally six bases such that
$\{
\left|00\right\rangle,\left|01\right\rangle,\left|02\right\rangle,
\left|10\right\rangle,\left|11\right\rangle,\left|12\right\rangle
\}.$ $A,B$ need to select rank$(\rho^C_{\Psi_{ABC}})$ basis from
the above set, so that the three cooperators can construct a
purely triple state. Let $n,m$ be integers, and $[a]$ denotes the
maximum integer not more than $a$. Due to the combinatorial
theory, the combination $\tbinom{n}{m}$ is monotonically
increasing when $m\leq [n/2]$ and monotonically decreasing when
$m\geq [n/2]$. So $\tbinom{n}{m}$ reaches its maximum when
$m=[n/2]$. We call this monotonicity quasi-combinatorial
character. Surprisingly, although not explicitly coinciding with
the combinations, the above numbers of entangled classes indeed
reflect this character. Another witness to this character can be
seen in the $2\times4\times N$ system, in which the
$2\times4\times4$ system can be entangled in infinitely many ways.
By virtue of the results in \cite{Chen}, there are orderly
$1,1,5,\infty,g,6,2,1$ classes of states in the $2\times4\times N$
space, $N=1,2,3,4,5,6,7,8,$ and we will provide the result of
$g=12$ later. Besides, there is always a unique class in $2\times
M\times 1$ and $2\times M\times 2M$ space respectively. So we
expect that there exists the quasi-combinatorial character in any
sequence of true $2\times M\times N$ systems, $N=1,2,...,2M.$ Due
to the experience in both this paper and \cite{Chen}, we can infer
that the classification of the $2\times M\times N$ states requires
an increasing amount of calculation when $N$ approaches $M$. At
last, the results with respect to the classification of
$2\times(M+3)\times(2M+3)$ and $2\times(M+4)\times(2M+4)$ states
are provided. For convenience, we list some existing results
appeared in \cite{Chen} in (i),(ii),(iii),
\\
(i) $2\times M\times2M,M\geq2:$
\begin{eqnarray*}
\left|\Upsilon_0\right\rangle\equiv\left|0\right\rangle\sum_{i=0}^{M-1}\left|ii\right\rangle
+\left|1\right\rangle\sum_{i=0}^{M-1}\left|i,i+M\right\rangle\in[0,0,\infty];
\end{eqnarray*}
(ii) $2\times(M+1)\times(2M+1),M\geq1:$
\begin{eqnarray*}
\left|\Upsilon_1\right\rangle&\equiv&\left|0,M,2M\right\rangle+\left|\Upsilon_0\right\rangle
\in[0,1,\infty];\\
\left|\Upsilon_2\right\rangle&\equiv&\left|0,M,2M\right\rangle+\left|1,M,M-1\right\rangle
+\left|\Upsilon_0\right\rangle\in[0,0,\infty];
\end{eqnarray*}
(iii) $2\times(M+2)\times(2M+2), M\geq2:$
\begin{eqnarray*}
\left|\Theta_0\right\rangle&\equiv&\left|1,M+1,2M+1\right\rangle+\left|\Upsilon_1\right\rangle\in[0,2,\infty];\\
\left|\Theta_1\right\rangle&\equiv&\left|0,M+1,2M+1\right\rangle+\left|\Upsilon_1\right\rangle\in[0,\infty,\infty];\\
\left|\Theta_2\right\rangle&\equiv&\left|1,M+1,2M+1\right\rangle+\left|\Upsilon_2\right\rangle\in[0,1,\infty];\\
\left|\Theta_3\right\rangle&\equiv&\left|0,M+1,2M+1\right\rangle+\left|1,M+1,2M\right\rangle
+\left|\Upsilon_1\right\rangle\\&\in&[0,1,\infty];\\
\left|\Theta_4\right\rangle&\equiv&\left|0,M+1,2M+1\right\rangle+\left|1,M+1,0\right\rangle
+\left|\Upsilon_2\right\rangle\\&\in&[0,0,\infty];\\
\left|\Theta_5\right\rangle&\equiv&\left|0,M+1,2M+1\right\rangle+\left|1,M+1,2M\right\rangle
+\left|\Upsilon_2\right\rangle\\&\in&[0,0,\infty].
\end{eqnarray*}
Based on these results we can carry on the new classification.
Recall that there exist the so-called invariance of
$\left|\Theta_0\right\rangle,\left|\Theta_4\right\rangle$ and
$\left|\Theta_5\right\rangle$, e.g., $\left|\Theta_0\right\rangle$
remains unchanged under ILO's
$\left|0\right\rangle_A\leftrightarrow\left|1\right\rangle_A$,
$\left|M+1\right\rangle_B\leftrightarrow\left|M\right\rangle_B$,
$\left|2M+1\right\rangle_C\leftrightarrow\left|2M\right\rangle_C$
and
$\left|i+M\right\rangle_C\leftrightarrow\left|i\right\rangle_C,i=0,...,M-1$.
One can find these invariances in \cite{Chen}, which will be
useful in the brief arguments for new outcomes. First we
investigate the case of $2\times(M+3)\times(2M+3)$ system. Due to
the method of LHRGM, the required condition is the above
$2\times(M+2)\times(2M+2)$ classes for the calculation of
$\left|\Omega_i\right\rangle,i=0,2,3$, while we have found out the
sole class $\left|\Gamma_{14}\right\rangle$ (see below) derived
from $\left|\Omega_1\right\rangle$ in (iv) in section II. As far
as the concrete process is concerned, by virtue of the techniques
in this paper and \cite{Chen} one can derive the classes
$\left|\Gamma_i\right\rangle,i=0,1,2,3,4,5,6,7,8$ from the family
of $\left|\Omega_0\right\rangle$, and
$\left|\Gamma_i\right\rangle,i=9,10,11,12,13$ from
$\left|\Omega_2\right\rangle$ and $\left|\Omega_3\right\rangle$,
which are two equivalent families when the low rank entangled
classes are
$\left|\Theta_0\right\rangle,\left|\Theta_4\right\rangle$ and
$\left|\Theta_5\right\rangle$ in terms of the invariances. In
addition, one can check that any pair of these classes are
inequivalent under SLOCC by the range criterion. Consequently, we
assert that the $2\times(M+3)\times(2M+3)$ system ($M\geq2$) can
be entangled in 15 ways in all under SLOCC.
\\(iv) $2\times(M+3)\times(2M+3), M\geq2:$ (For the case of $M=1$,
$\left|\Gamma_7\right\rangle$, $\left|\Gamma_{11}\right\rangle$,
$\left|\Gamma_{13}\right\rangle$ disappear due to the absence of
$\left|\Theta_4\right\rangle_{2\times3\times4}$)
\begin{eqnarray*}
\left|\Gamma_0\right\rangle&\equiv&\left|1,M+2,2M+2\right\rangle+\left|\Theta_0\right\rangle
\in[0,\infty,\infty];\\
\left|\Gamma_1\right\rangle&\equiv&(\left|0\right\rangle+\left|1\right\rangle)
\left|M+2,2M+2\right\rangle+\left|\Theta_0\right\rangle
\in[0,3,\infty];\\
\left|\Gamma_2\right\rangle&\equiv&\left|0,M+2,2M+2\right\rangle+\left|\Theta_1\right\rangle
\in[0,\infty,\infty];\\
\left|\Gamma_3\right\rangle&\equiv&\left|1,M+2,2M+2\right\rangle+\left|\Theta_2\right\rangle
\in[0,\infty,\infty];\\
\left|\Gamma_4\right\rangle&\equiv&\left|0,M+2,2M+2\right\rangle+\left|\Theta_2\right\rangle
\in[0,2,\infty];\\
\left|\Gamma_5\right\rangle&\equiv&\left|1,M+2,2M+2\right\rangle+\left|\Theta_3\right\rangle
\in[0,2,\infty];\\
\left|\Gamma_6\right\rangle&\equiv&\left|0,M+2,2M+2\right\rangle+\left|\Theta_3\right\rangle
\in[0,\infty,\infty];\\
\left|\Gamma_7\right\rangle&\equiv&\left|1,M+2,2M+2\right\rangle+\left|\Theta_4\right\rangle
\in[0,1,\infty];\\
\left|\Gamma_8\right\rangle&\equiv&\left|1,M+2,2M+2\right\rangle+\left|\Theta_5\right\rangle
\in[0,1,\infty];\\
\left|\Gamma_9\right\rangle&\equiv&\left|1,M+2,2M+2\right\rangle+\left|0,M+2,2M+1\right\rangle
+\left|\Theta_2\right\rangle\\&\in&[0,1,\infty];\\
\left|\Gamma_{10}\right\rangle&\equiv&\left|0,M+2,2M+2\right\rangle
+\left|1,M+2,2M+1\right\rangle+\left|\Theta_3\right\rangle\\
&\in&[0,1,\infty];\\
\left|\Gamma_{11}\right\rangle&\equiv&\left|1,M+2,2M+2\right\rangle
+\left|0,M+2,M+1\right\rangle+\left|\Theta_4\right\rangle\\
&\in&[0,0,\infty];\\
\left|\Gamma_{12}\right\rangle&\equiv&\left|1,M+2,2M+2\right\rangle
+\left|0,M+2,M\right\rangle+\left|\Theta_5\right\rangle\\
&\in&[0,0,\infty];\\
\left|\Gamma_{13}\right\rangle&\equiv&\left|0,M+2,2M+2\right\rangle
+\left|1,M+2,2M+1\right\rangle+\left|\Theta_5\right\rangle\\
&\in&[0,0,\infty];\\
\left|\Gamma_{14}\right\rangle&\equiv&\sum_{i=0}^{M-1}(\left|0,M+2-i,2M+2-2i\right\rangle\\&+&
\left|1,M+2-i,2M+1-2i\right\rangle)+\left|\varphi_2\right\rangle\in[0,\infty,\infty].\\
\end{eqnarray*}
Based on the above results we can further classify the
entanglement of $2\times(M+4)\times(2M+4)$ system. Notice the
calculation of $\left|\Omega_1\right\rangle$ requires the
classification of the $2\times4\times4$ states, which has been
given in section III and only the class
$\left|\Lambda_{36}\right\rangle$ (see below) is derived. By using
of the LHRGM one can obtain the classes
$\left|\Lambda_i\right\rangle,i\in[0,23]$ from the family of
$\left|\Omega_0\right\rangle$, and the classes
$\left|\Lambda_i\right\rangle,i\in[24,35]$ from the family of
$\left|\Omega_2\right\rangle$ and $\left|\Omega_3\right\rangle$.
In particular, $\left|\Lambda_3\right\rangle$ contains a parameter
which cannot be removed, and thus the $2\times(M+4)\times(2M+4)$
states are also a family of infinitely many classes. One can
analyze the family of $\left|\Lambda_3\right\rangle$ similar to
the case of $\left|\Phi_3\right\rangle_x$ in $2\times4\times4$
space, since there are only four product states in
$R(\rho^{AC}_{\Phi_3x})$, and the possible transformations indeed
make a term in the initial state into another term in the final
state. We thus get the results of a new family containing
infinitely many classes by the existing techniques, although the
calculation here is more involved than the former cases.

(v) $2\times(M+4)\times(2M+4),M\geq2:$ (For the case of $M=1$,
$\left|\Lambda_i\right\rangle,i=12,13,20,22,25,28,29,31,32$
disappear and an added class is
$\left|045\right\rangle+\left|142\right\rangle+\left|\Gamma_9\right\rangle^{M=1}$)
\begin{eqnarray*}
\left|\Lambda_0\right\rangle&\equiv&\left|1,M+3,2M+3\right\rangle+\left|\Gamma_0\right\rangle
\in[0,\infty,\infty];\\
\left|\Lambda_1\right\rangle&\equiv&\left|0,M+3,2M+3\right\rangle+\left|\Gamma_0\right\rangle
\in[0,\infty,\infty];\\
\left|\Lambda_2\right\rangle&\equiv&(\left|0\right\rangle+\left|1\right\rangle)\left|M+3,2M+3\right\rangle
+\left|\Gamma_0\right\rangle\in[0,\infty,\infty];\\
\left|\Lambda_3\right\rangle&\equiv&(\left|0\right\rangle+x\left|1\right\rangle)\left|M+3,2M+3\right\rangle
+\left|\Gamma_1\right\rangle\\&\in&[0,4,\infty],x\neq0,1;\\
\left|\Lambda_4\right\rangle&\equiv&\left|0,M+3,2M+3\right\rangle+\left|\Gamma_2\right\rangle
\in[0,\infty,\infty];\\
\left|\Lambda_5\right\rangle&\equiv&\left|1,M+3,2M+3\right\rangle+\left|\Gamma_3\right\rangle
\in[0,\infty,\infty];\\
\left|\Lambda_6\right\rangle&\equiv&\left|0,M+3,2M+3\right\rangle+\left|\Gamma_3\right\rangle
\in[0,\infty,\infty];\\
\left|\Lambda_7\right\rangle&\equiv&(\left|0\right\rangle+\left|1\right\rangle)\left|M+3,2M+3\right\rangle
+\left|\Gamma_4\right\rangle\in[0,3,\infty];\\
\left|\Lambda_8\right\rangle&\equiv&\left|1,M+3,2M+3\right\rangle+\left|\Gamma_5\right\rangle
\in[0,\infty,\infty];\\
\left|\Lambda_9\right\rangle&\equiv&\left|0,M+3,2M+3\right\rangle+\left|\Gamma_5\right\rangle
\in[0,\infty,\infty];\\
\left|\Lambda_{10}\right\rangle&\equiv&(\left|0\right\rangle+\left|1\right\rangle)\left|M+3,2M+3\right\rangle
+\left|\Gamma_5\right\rangle\in[0,3,\infty];\\
\left|\Lambda_{11}\right\rangle&\equiv&\left|0,M+3,2M+3\right\rangle+\left|\Gamma_6\right\rangle
\in[0,\infty,\infty];\\
\left|\Lambda_{12}\right\rangle&\equiv&\left|1,M+3,2M+3\right\rangle+\left|\Gamma_7\right\rangle
\in[0,\infty,\infty];\\
\left|\Lambda_{13}\right\rangle&\equiv&\left|0,M+3,2M+3\right\rangle+\left|\Gamma_7\right\rangle
\in[0,2,\infty];\\
\left|\Lambda_{14}\right\rangle&\equiv&\left|1,M+3,2M+3\right\rangle+\left|\Gamma_8\right\rangle
\in[0,\infty,\infty];\\
\left|\Lambda_{15}\right\rangle&\equiv&\left|0,M+3,2M+3\right\rangle+\left|\Gamma_8\right\rangle
\in[0,2,\infty];\\
\left|\Lambda_{16}\right\rangle&\equiv&\left|1,M+3,2M+3\right\rangle+\left|\Gamma_9\right\rangle
\in[0,\infty,\infty];\\
\left|\Lambda_{17}\right\rangle&\equiv&\left|0,M+3,2M+3\right\rangle+\left|\Gamma_9\right\rangle
\in[0,2,\infty];\\
\left|\Lambda_{18}\right\rangle&\equiv&\left|1,M+3,2M+3\right\rangle+\left|\Gamma_{10}\right\rangle
\in[0,2,\infty];\\
\left|\Lambda_{19}\right\rangle&\equiv&\left|0,M+3,2M+3\right\rangle+\left|\Gamma_{10}\right\rangle
\in[0,\infty,\infty];\\
\left|\Lambda_{20}\right\rangle&\equiv&\left|1,M+3,2M+3\right\rangle+\left|\Gamma_{11}\right\rangle
\in[0,1,\infty];\\
\left|\Lambda_{21}\right\rangle&\equiv&\left|1,M+3,2M+3\right\rangle+\left|\Gamma_{12}\right\rangle
\in[0,1,\infty];\\
\left|\Lambda_{22}\right\rangle&\equiv&\left|1,M+3,2M+3\right\rangle+\left|\Gamma_{13}\right\rangle
\in[0,1,\infty];\\
\left|\Lambda_{23}\right\rangle&\equiv&\left|1,M+3,2M+3\right\rangle+\left|\Gamma_{14}\right\rangle
\in[0,\infty,\infty];\\
\left|\Lambda_{24}\right\rangle&\equiv&\left|1,M+3,2M+3\right\rangle+\left|0,M+3,2M+2\right\rangle
\\&+&\left|\Gamma_5\right\rangle\in[0,2,\infty];\\
\left|\Lambda_{25}\right\rangle&\equiv&\left|1,M+3,2M+3\right\rangle+\left|0,M+3,2M+2\right\rangle
\\&+&\left|\Gamma_7\right\rangle\in[0,1,\infty];\\
\left|\Lambda_{26}\right\rangle&\equiv&\left|1,M+3,2M+3\right\rangle+\left|0,M+3,2M+2\right\rangle
\\&+&\left|\Gamma_8\right\rangle\in[0,1,\infty];\\
\left|\Lambda_{27}\right\rangle&\equiv&\left|1,M+3,2M+3\right\rangle+\left|0,M+3,2M+2\right\rangle
\\&+&\left|\Gamma_9\right\rangle\in[0,1,\infty];\\
\left|\Lambda_{28}\right\rangle&\equiv&\left|1,M+3,2M+3\right\rangle+\left|0,M+3,M+2\right\rangle
\\&+&\left|\Gamma_{11}\right\rangle\in[0,0,\infty];\\
\left|\Lambda_{29}\right\rangle&\equiv&\left|1,M+3,2M+3\right\rangle+\left|0,M+3,2M+2\right\rangle
\\&+&\left|\Gamma_{11}\right\rangle\in[0,0,\infty];\\
\end{eqnarray*}
\begin{eqnarray*}
\left|\Lambda_{30}\right\rangle&\equiv&\left|1,M+3,2M+3\right\rangle+\left|0,M+3,2M+2\right\rangle
\\&+&\left|\Gamma_{12}\right\rangle\in[0,0,\infty];\\
\left|\Lambda_{31}\right\rangle&\equiv&\left|0,M+3,2M+3\right\rangle+\left|1,M+3,0\right\rangle
\\&+&\left|\Gamma_{13}\right\rangle\in[0,0,\infty];\\
\left|\Lambda_{32}\right\rangle&\equiv&\left|0,M+3,2M+3\right\rangle+\left|1,M+3,2M+2\right\rangle
\\&+&\left|\Gamma_{13}\right\rangle\in[0,0,\infty];\\
\left|\Lambda_{33}\right\rangle&\equiv&\left|1,M+3,2M+3\right\rangle+\left|0,M+3,2M+1\right\rangle
\\&+&\left|\Gamma_{14}\right\rangle\in[0,\infty,\infty];\\
\left|\Lambda_{34}\right\rangle&\equiv&\left|0,M+3,2M+3\right\rangle+\left|1,M+3,2M+2\right\rangle
\\&+&\left|\Gamma_6\right\rangle\in[0,\infty,\infty];\\
\left|\Lambda_{35}\right\rangle&\equiv&\left|0,M+3,2M+3\right\rangle+\left|1,M+3,2M+2\right\rangle
\\&+&\left|\Gamma_{10}\right\rangle\in[0,1,\infty];\\
\left|\Lambda_{36}\right\rangle&\equiv&\sum_{i=0}^{M-1}(\left|0,M+3-i,2M+3-2i\right\rangle\\&+&
\left|1,M+3-i,2M+2-2i\right\rangle)\\&+&\left|\Phi_{15}\right\rangle\in[0,\infty,\infty].\\
\end{eqnarray*}
All these states are incomparable under SLOCC by the range
criterion. The laconic structures show the regular generation of
the higher level classes derived from the low-level classes.

\section{conclusions}

In this paper, we mainly classified the entangled states of
$2\times4\times4$ system, which is the simplest one containing
infinitely many classes. So the range criterion is indeed useful
for the classification of generic entanglement which usually
contains parameters, and the results of the $2\times4\times4$
states helps explore the structure of 5-qubit entanglement. It
turned out that the range criterion efficiently operates for the
discrimination of multipartite entangled states, and in principle
one can classify any family of true $2\times M\times N$
entanglement by virtue of the LHRGM. Finally we have managed to
obtain the classification of $2\times(M+3)\times(2M+3)$ and
$2\times(M+4)\times(2M+4)$ states. These results are helpful to
the further study in this aspect.

There are two main points we can anticipate from this paper.
First, it is worth considering the asymmetry arising from the
$2\times4\times4$ entanglement. This phenomenon should be an
important feature of the $2\times M\times M$ entanglement, whose
classification is more complicated than the $2\times M\times N$
cases where $M\neq N$. In addition, the study of this property is
propitious for the more generic cases such as the 3-qutrit
entanglement. Second, we are going to apply the range criterion to
the classification of multiqubit system, which remains a difficult
and important problem in QIT. However, since there exist more
symmetry in the general multipartite system, other techniques may
be required for the further exploration.

The work was partly supported by the NNSF of China Grant
No.90503009 and 973 Program Grant No.2005CB724508.


\begin{references}
\bibitem{notation1} As defined in section III, the states
$\left|\Phi_{11}\right\rangle$ and $\left|\Phi_{15}\right\rangle$
are two essential classes, and it is easy to find that
$\left|\Phi_{11}\right\rangle_{ABC}\sim\left|\Phi_{15}\right\rangle_{ACB}$
by some ILO's. We infer that this property also exists in the
systems with higher dimension, and this is an interesting
phenomenon requiring a further investigation.
\end{references}
\end{document}